\documentclass{scrartcl}
\usepackage[utf8]{inputenc}
\usepackage{comment}
\usepackage{amsthm}
\usepackage{amsmath}
\usepackage{amssymb}
\usepackage{array}
\usepackage{tabularx}
\usepackage{booktabs}

\usepackage{subcaption}

\usepackage[authoryear,round,numbers]{natbib}

\title{Optimal Portfolio Execution in a Regime-switching Market with Non-linear Impact Costs: Combining Dynamic Program and Neural Network}

\author{Xiaoyue Li \and John M. Mulvey}
\date{August 15, 2021}

\usepackage{graphicx}

\begin{document}

\maketitle

\section{Introduction}
The optimal execution of large amounts of portfolios have long been an important problem for institutional investors. When liquidating or acquiring a large position over a given horizon, each order the investor puts would move the market price in the unfavorable direction. If the investor implements the position rapidly, she consumes deeply into the order book and therefore cannot trade at desired price. On the other hand, if she executes at a slower pace, the underlying assets are exposed to market volatility, resulting in a loose risk control of her execution price. \cite{Almgren1999} provide a detailed description of the trade-off between expected liquidized value and its standard deviation that the trade faces.\\
\\
\cite{bertsimas1998} employ dynamic program to derive the optimal execution strategy of one security that minimizes the expected trading cost during a fixed horizon, when the market impact cost functions are given. Under linear price impacts, they extend the model and address the same problem when there are multiple underlying risky assets to be executed (\cite{bertsimas1999}). \cite{almgren2000} also consider an impact with temporary and permanent parts, where the former only affects the current trade, and the latter moves the asset price permanently. For a linear cost model, they construct the mean-variance efficient frontier and address the trade-off between reward and risk. Later, \cite{Almgren2003} extends the model to allow for nonlinear increasing impact costs and increasing variance of realized price movement with respect to trading speed, and solves the exact solution when the impact cost is a power law function. Since then, plentiful studies have been made on market impact functions as well as on trading strategies of optimal execution problems, analytically or numerically.\\
\\
A line of research is based on order books that describe supply and demand dynamics. \cite{obizhaeva2005} assume a block-shaped limit order book that recovers to its steady state at an exponential rate, and minimize (maximize) the expected total cost (gain) of asset purchase (sale). \cite{alfonsi2010} extend the model to allow for a general shape of the limit order book. They analyze the optimal strategy in discrete time under two models for price resilience: one with exponential recovery of limit order book, the other with exponential recovery of the ask-bid spread. \cite{cartea2014} present a strategy to trade a large position, taking advantage of both limit and market orders. \cite{alfonsi2015} offer closed-form strategy under a linear price impact model with block-shaped order book where the orders come at a Hawkes process. \cite{siu2019} provide an optimal strategy for market order placement in a limit order book market, in order to minimize the expected cost of acquiring a position. Their limit order book is assumed to be block-shaped, but the rate of recovering after each trade is relaxed to be Markovian regime-switching. They conclude that an investor shall trade more aggressive when the limit order book switches from a low to a high resilience state, and vise versa. \\
\\
Some studies take the price impact functions as given, and build analytical or numerical methods to solve for the optimal trading schedule. \cite{forsyth2010} models the market impact costs as a function of trading speed, and formulates a mean-variance optimization problem that can be embedded in a linear-quadratic stochastic control. Based on the Almgren-Chriss framework, \cite{gatheral2010} assume the price follows a geometric Brownian motion, and find the closed-form solution using Hamilton-Jacobi-Bellman equations, whose robustness with respect to model misspecification is shown by \cite{Schied2013}. \cite{moazeni2010} consider a linear price impact with a temporary component $\Omega$ and a permanent component $\Gamma$, and find that the optimal execution strategy is characterized by $\frac{1}{\tau}(\Omega+\Omega^T)-\Gamma$ where $\tau$ is the time window between trades. They present the mean-variance efficient frontier and test the sensitivity of strategies to price impact parameters. \cite{kato2014} assumes a continuous-time market model, and solves the stochastic control problem for the trading schedule. \cite{curato2014} introduce a fully numerical method based on homotopy analysis, in a market where impact propagator decays exponentially asymptotically. \cite{kalsi2019} build a general framework requiring a geometric rough price process and a continuous price impact function, and employ signature method to find approximated solutions to the optimal execution problem.\\
\\
With the development of computational power, advanced machine learning methods are applied to financial problems including optimal execution, portfolio management, Greek estimations, etc. \cite{ning2020} utilize neural networks trained with experience replay and double Q-learning to tackle the optimal execution problem. They hire a model-free approach, and assume no market impact costs when training the model. Their algorithm is tested on nine different equities, and compared to the benchmark strategy that trades equal amount at each time period. The machine learning method outperforms the benchmark in most of the tested equities. In October 2020, Royal Bank of Canada launches its AI-based electronic trading platform, Aiden, who applies a model-free reinforcement learning with simulated market environment based on historical trading data. It takes hundreds of market and self-aware features and learns to optimize a trading objective. To improve stability, Aiden employs actor-critic approach (\cite{sutton2018}) and proximal policy optimization (\cite{schulman2017}). \\
\\
Our focus of this paper is a numerical method for the optimal execution problem of a portfolio with multiple assets, based on a generic price impact framework. We assume trades affect the market with a temporary component and a permanent one. However, we do not rule the exact form of either. For illustration reasons, we present results with a quadratic convex market impact functions, but the impact can be any function of the trading amount. The optimal execution of a portfolio is generally a much harder problem than that of one asset. When correlated assets are included in the portfolio, trading one asset may lead to the undesirable price movement of another, which complicates the trading problem. In this paper, we break down the portfolios to approximated orthogonal portfolios. For each of the orthogonal portfolios, a dynamic program is hired to find the optimal execution strategy. To correct error terms brought by approximating parameters for each orthogonal portfolio, we train a neural network on the original problem to learn the optimal trading schedule of the original portfolio. For simplicity, we will focus on sale of large portfolios. Our method is capable of optimizing the CRRA (constant relative risk aversion) utility of the terminal wealth, as well as solving the mean-variance formulation. Comparing to the benchmark strategy that trades equal amount at each period, our combined method achieves significantly higher expected terminal wealth in the CRRA utility case, and attains moderate higher expected wealth with notable improvement in risk control with the mean-variance formulation. Another possible benchmark is to compare the execution price to the value-weighted average price (VWAP), which is known only after the whole trading period. To compare with VWAP, an important aspect is to predict the volume pattern, which is not the focus of this paper. Therefore, we choose to use the benchmark of equal trades per period, and optimize on absolute objectives such as CRRA and mean-variance.\\
\\
The main contributions of this work is three-fold. First, the model we consider is a generic one that does not require the specific form of market impact costs. By utilizing a numerical method, our model can solve for a selling strategy for any continuous impact cost. In addition, we allow regime switching in the model, to provide more flexibility and better mimic the real market. Second, we introduce a method to break down an optimal execution problem on a portfolio to subproblems of trading single securities. The trading schedules suggested by the approximated orthogonal portfolio approach, though suboptimal, beats the benchmark of equal trades per period in many cases, and provides an advanced starting point for further machine learning improvement. Last, we propose a combined method that links a neural network with solutions from the approximated orthogonal portfolio approach. The result outperforms the benchmark, and is tested to be robust. We plot the mean-variance efficient, and discuss the merit and demerit of CRRA versus mean-variance objective functions. Whereas CRRA objectives usually offer high expected terminal wealth by selling the portfolio, mean-variance objectives provide a better control between returns and risks.\\
\\
The rest of the paper is organized as follows. In Section 2, we provide a full description of the optimal execution model of portfolios in a regime-switching market with impact costs. Section 3 introduces the approach we apply to tackle the problem of interest. Numerical examples of CRRA objective with various complexity appear in Section 4. We present examples of mean-variance objective function, and provide the mean-variance efficient frontier in Section 5. Section 6 discusses the running time of our combined method. Section 7 concludes.\\
\\
\section{The model} 
We focus on the following problem in this paper. Suppose a trader needs to complete a large trade of purchase or sale of a portfolio by some time $T$, with underlying securities $\mathbb{J}={1,...,n}$. The goal is to spend as little cash to accomplish the purchase, or to gain as much by selling. The horizon is discretized into $T$ periods, $t=1,2,...,T$ where the trader may place the order at the beginning of each period.\\
\\
There are $m\geq 1$ regimes in the market, depending on which the asset price dynamics and/or market impact costs may vary. For example, there may be regimes under which the price grows steadily, and under which the transaction costs are higher due to illiquidity. We assume the returns in each period follows a normal distribution whose parameters only depend on underlying regime. Regime switching happens only at the end of each period, and follows a Markov process. We assume the regime is directly observable to the trader, but she cannot foresee the future regime.\\
\\
Trading comes with costs. Two types of costs are considered in our model: a temporary transaction cost $tr_{i}^a$, and a permanent transaction cost $tr_{i}^b$, where $i$ is the underlying regime. Both types of transaction costs are modelled with functions of the amount of assets traded. Consider a typical order book. There is limited stack at one's favorable price. To trade a large amount, one needs to consume deep into the order book, and therefore trades at a price deviating from the best available price. We call it the temporary transaction cost. After the current order is complete, the mid price is pushed against the trader, and usually will not recover immediately. The impact of the order on the future price is called the permanent transaction cost. Typically, a trader does not complete the order at once; on the other hand, she may accomplish the trade in smaller chunks so as to reduce the impact of transaction costs.\\
\\
The cases of purchase and sale of a portfolio are symmetric. For simplicity, we only discuss the selling case in this paper. We will consider two families of objective functions: 
\begin{itemize}
    \item expected CRRA (constant relative risk aversion) utility, where we maximize the expected utility of terminal wealth, $\mathbb{E}[U(w)]$, where $U(w) = 
\begin{cases}
\frac{w^\gamma}{\gamma} & \text{ if $\gamma\neq 0$ }\\
ln(w) & \text{ if $\gamma= 0$ }
\end{cases}$ for risk aversion coefficient $\gamma$; and 
\item mean-variance utility, where we optimize on $\mathbb{E}[w]-\lambda Var(w)$ with mean-variance coefficient $\lambda$.
\end{itemize}
\noindent
The method we discuss in this paper is suitable for any impact costs that are functions of the amount of assets traded. For illustration we limit ourselves to examples with both temporary transaction costs and permanent transaction costs being second-order (convex) functions.
In particular, in each regime $i$, temporary transaction costs are represented by two matrices $tr_i^{a,\alpha}=
\begin{bmatrix}
tr_{i,(1,1)}^{a,\alpha} & tr_{i,(1,2)}^{a,\alpha} & ... & tr_{i,(1,n)}^{a,\alpha}\\
tr_{i,(2,1)}^{a,\alpha} & tr_{i,(2,2)}^{a,\alpha} & ... & tr_{i,(2,n)}^{a,\alpha}\\
... & ... & ... & ...\\
tr_{i,(n,1)}^{a,\alpha} & tr_{i,(n,2)}^{a,\alpha} & ... & tr_{i,(n,n)}^{a,\alpha}\\
\end{bmatrix}$ and $tr_i^{a,\beta}=
\begin{bmatrix}
tr_{i,(1,1)}^{a,\beta} & tr_{i,(1,2)}^{a,\beta} & ... & tr_{i,(1,n)}^{a,\beta}\\
tr_{i,(2,1)}^{a,\beta} & tr_{i,(2,2)}^{a,\beta} & ... & tr_{i,(2,n)}^{a,\beta}\\
... & ... & ... & ...\\
tr_{i,(n,1)}^{a,\beta} & tr_{i,(n,2)}^{a,\beta} & ... & tr_{i,(n,n)}^{a,\beta}\\
\end{bmatrix}$, where $n$ is the number of risky assets involved. When $x_1$ chunks of asset 1, $x_2$ chunks of asset 2, ..., and $x_n$ chunks of asset n are sold simultaneously at some point, the temporary transaction costs incurred on asset $k$ is $\Sigma_{j=1}^n tr_{i,(k,j)}^{a,\alpha}x_j + tr_{i,(k,j)}^{a,\beta}x_j^2$. I.e., the temporary transaction costs to all assets can be written as a column vector $tr_i^{a,\alpha}\mathbf{x} + tr_i^{a,\beta}(\mathbf{x}\odot\mathbf{x})$ where $\mathbf{x} = [x_1, x_2, ..., x_n]^T$ and $\odot$ denotes element-wise multiplication. Similarly, permanent transaction costs in regime $i$ is indicated by two matrices $tr_i^{b,\alpha}$ and $tr_i^{b,\beta}$ which includes all first order and second order terms, respectively. The permanent transaction costs under regime $i$ is found by $tr_i^{b,\alpha}\mathbf{x} + tr_i^{b,\beta}(\mathbf{x}\odot\mathbf{x})$ with same notations as before.\\

\section{Methodology}
\subsection{Dynamic program}
Dynamic program has brought profound influence on sequential decision problems since \cite{bellman}. It takes advantage of backward recursion, and solves an optimization problem by breaking it down to simpler subproblems and reusing the solutions to these subproblems. \\
\\
A natural algorithm for multi-stage problems, we tackle the problem of selling one security with dynamic program. The goal is set to maximize the expected utility of cash at the horizon with risk aversion coefficient $\gamma$.
In our setting, the solution is obvious at the horizon $t=T$: to sell all remaining chunks. We then take advantage of the information at $t=T$ to solve the strategy at $t=T-1$, and so on, until the original problem is solved.\\
\\
\textbf{State space.} The state space consists of all the components that are not under the agent's direct control. The state may be affected by the agent's previous decisions, but the agent cannot take complete control of it. The agent is able to observe the state, either partially or fully, and make decision based on the observation. For the trading problem, we define the state to be a tuple $(i, C, S, t)$, where $i\in\{1,...,m\}$ is the current regime, $C$ is the amount of realized cash normalized by the current asset price, $S$ is the number of chunks yet to be redeemed, and $t$ is the time. The state space is $\mathbf{S} = \{(i, C, S, t): i\in\{1,...,m\}, C\geq0, 0\leq S\leq S_0, 0\leq t\leq T\}$ for initial number of chunks $S_0$. The initial state is $(0,S_0,0)$. CRRA utility allows us to define the state without current asset price, since it can be easily scaled with $U(\alpha w) = \alpha^\gamma U(w)$. \\ 
\\
\textbf{Action space.} The action space is a set, $\mathbf{A}$, of all actions that the agent may take. In our setting, given a state $(i_t, C_t, S_t, t)$, possible actions to take are to sell $x_t$ chunks for $x_t\in\{0,1,...,S\}$.\\
\\
\textbf{Value function.} A value function $V:\mathbf{S}\rightarrow \mathbb{R}$ maps states to real numbers. It evaluates how well an agent may end up given that the agent is currently in some state $s\in\mathbf{S}$. Here, we define the value function to be the expected terminal utility following the optimal strategy from the current state to the horizon (cost-to-go), and assuming that the current wealth is \$1.
$$ V(i_t, C_t,S_t,t) = \underset{\{x_t,...,x_T\}}{\text{Maximize}}\mathbf{E}[U(W_T)|(i_t,C_t,S_t, t), p_t=1]\hspace{1cm}  \forall (i_t,C_t,S_t,t)\in\mathbf{S},$$
where $p_t$ is the price of the asset at time $t$. After selling $x_t$ chunks at time $t$, the number of remaining chunks to sell becomes $S_t-x_t$, and the trader gains $x_t p_t(1-tr_{i_t}^a(x_t))$. The value function satisfies the Bellman equation:
$$V(i_t, C_t,S_t,t) = \underset{\{x_t\}}{\text{Maximize}}\mathbf{E}[\underset{j}{\Sigma}\mathbb{P}(i_{t+1}=j|i_t)p_{t+1}^\gamma V(j, \frac{C_t+x_t(1-tr_{i_t}^a(x_t))}{p_{t+1}}, S_t-x_t, t+1)].$$
\\
Decisions at time $t$ depend only on the information up to time $t$ and does not avoid the non-anticipativity constraints.

\subsection{Orthogonal portfolios}
Dynamic program suffers from the curse of dimensionality, the phenomenon that running time grows exponentially when the complexity of the problem increases. In our problem, dynamic program presents an elegant solution for the case of the target portfolio containing one security. However, as the number of securities of interest increases, one requires more information in the state space for a decision to be made. For example, the relationship of the prices of these securities must be included to fully describe the state, and the relationship vector would have dimension $n-1$. The number of securities empirically jeopardizes the effectiveness of dynamic program if we apply the method directly. In this subsection, we introduce a way to reduce the dimension of multi-asset optimal execution problem.\\
\\
We use the term \textit{orthogonal portfolios} to describe several portfolios whose returns are orthogonal to each other.
Intuitively, if a portfolio has zero correlation with another, trading one of them should not impact the price of the other. This phenomenon indicates that the permanent transaction costs on one of the orthogonal portfolios is approximately zero when another is traded. In particular, \cite{mastromatteo2017} propose the EigenLiquidity Model with which they model the cross-impact matrix from the eigenvectors of the correlation matrix. They argue that it is reasonable to assume that the correlation matrix and the cross-impact matrix share the same set of eigenvectors, which implies that the portfolios corresponding to these eigenvectors shall be relatively uncorrelated to each other. Employing this idea, we create $n$ \textit{approximately} orthogonal portfolios in this optimal execution problem, by using the number of average chunks to be sold in each period. For each of the approximately orthogonal portfolios, we optimize the trading schedule with dynamic program. In the end, all trading schedules are converted back to the original assets.\\

\subsection{Neural networks}
Based on the approximate orthogonal portfolios, we learned a strategy that works well. Yet, errors are introduced when we approximate the transaction costs for each orthogonal portfolio, leading to suboptimal solutions. We seek to improve the performance on top of the results from the approximation method. \\
\\
We will employ a fully-connected feedforward neural network that contains one hidden layer. The input layer includes all information that an agent needs in order to make a selling decision: the current period $t$, current regime, amounts left to be sold for all assets, current prices of all assets, and the current accumulated cash. The output layer consists of $n$ neurons, which are the amounts to be sold for each of the risky assets. The computational graph of the neural network appears in Figure \ref{fig:nn}.\\
\begin{figure}[ht!]
\center
\includegraphics[width=0.6\linewidth]{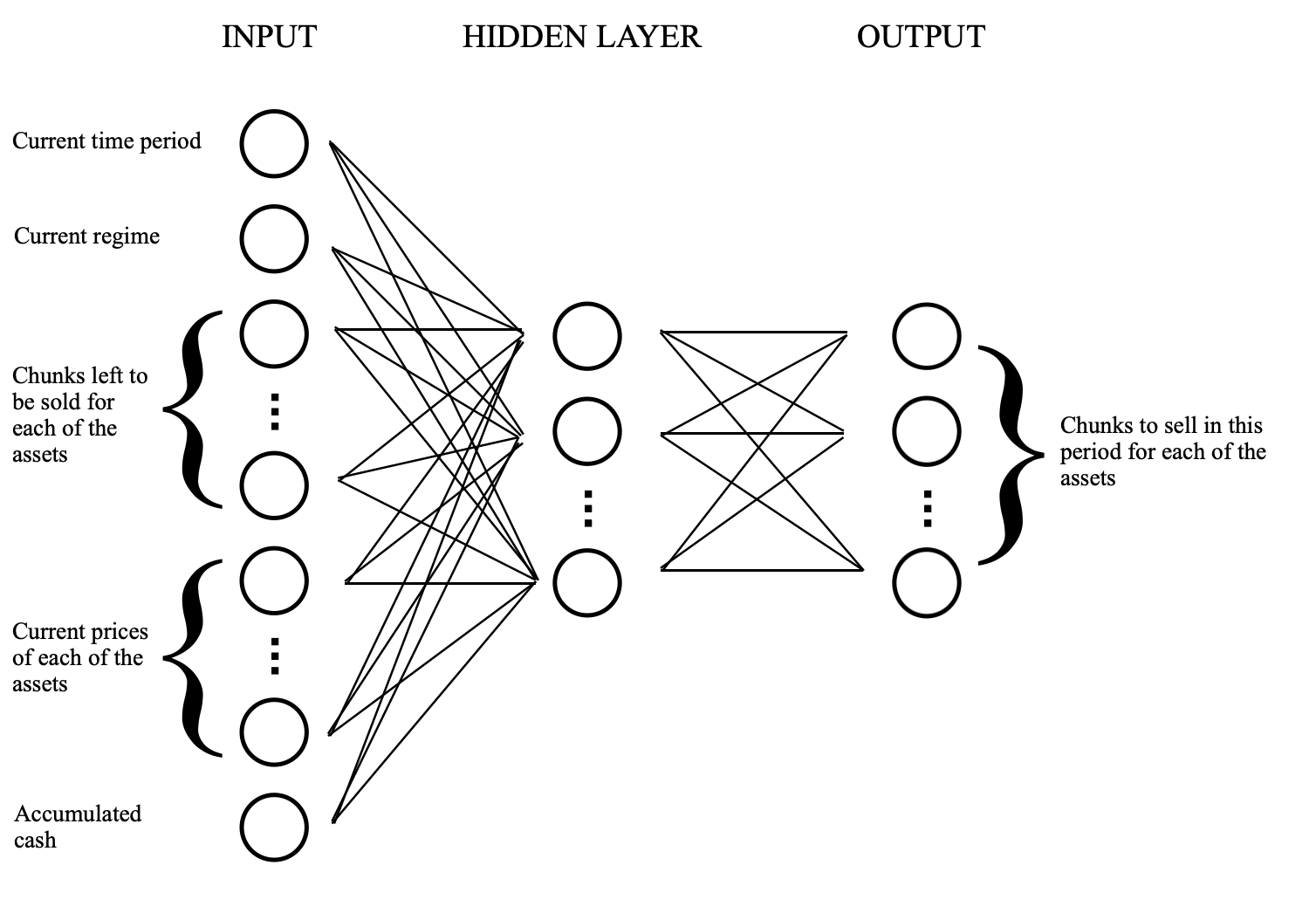} 
\caption{Computational graph of the neural network.}
\label{fig:nn}
\end{figure}\\
\\
First, we pre-train the weights using the strategy we learned from approximated orthogonal portfolios. Recall that the world-class computer program AlphaGo starts its training with learning human players' moves (\cite{alphago}). This is an analogue where we perform a supervised learning where the objective is to minimize the sum of squares of the difference between output layer and the strategy learned with approximate orthogonal portfolios. The goal of this step is to ensure the outputs will be relatively close to the theoretical optimal strategy, and thus would not be trapped at other local optima. 
We record these trained weights, and apply a small random turbulence to the pre-trained weights. They will serve as starting values, or initial weights, as we proceed to the next step.\\
\\
Lastly, we feed the trading model to a neural network with the same structure to optimize directly on the expected utility of terminal wealth. The neural network is trained with Adam Optimizer based on gradient method, with leaky ReLU activation function.\\

\section{Example with multiple risky assets}
Recall that the trading problem with one asset is solved directly with dynamic programs. Given the limited space, we present examples on one-asset case with sensitivity analysis in the appendix. In this section, we focus on computational results of optimal execution problems on multiple assets to illustrate the power of dimension reduction and neural networks.

\subsection{An example with 3 assets}\label{example:3asset}
Suppose there are three risky assets to be liquidized over 10 periods, each 20 chunks. The first and second order term of the temporary transaction costs in Regime 1 are 
$tr_1^{a,\alpha}= 0.002*
\begin{bmatrix}
1& 0.2& 0.2\\
0.2& 1& 0.2\\
0.2& 0.2& 1\\
\end{bmatrix}$ and $tr_1^{a,\beta}=10^{-4}*
\begin{bmatrix}
1& 0.2& 0.2\\
0.2& 1& 0.2\\
0.2& 0.2& 1\\
\end{bmatrix}$, respectively; those in Regime 2 are
$tr_2^{a,\alpha}= 0.002*
\begin{bmatrix}
1& 0.2& 0.2\\
0.2& 2& 0.2\\
0.2& 0.2& 2\\
\end{bmatrix}$ and $tr_2^{a,\beta}=10^{-4}*
\begin{bmatrix}
1& 0.3& 0.3\\
0.3& 2& 0.3\\
0.3& 0.2& 3\\
\end{bmatrix}$, respectively. \\
\\
The first and second order of the permanent transaction costs under Regime 1 are
$tr_1^{b,\alpha}= 10^{-4}*
\begin{bmatrix}
1& 0.1& 0.1\\
0.2& 1& 0.1\\
0.1& 0.1& 1\\
\end{bmatrix}$ and $tr_1^{b,\beta}=10^{-4}*
\begin{bmatrix}
1& 0.1& 0.1\\
0.2& 1& 0.1\\
0.1& 0.1& 1\\
\end{bmatrix}$, respectively; those in Regime 2 are
$tr_2^{b,\alpha}= 10^{-4}*
\begin{bmatrix}
2& 0.8& 0.4\\
0.4& 4& 0.4\\
0.4& 0.4& 2\\
\end{bmatrix}$ and $tr_2^{b,\beta}=10^{-4}*
\begin{bmatrix}
3& 0.8& 0.4\\
0.4& 4& 0.4\\
0.4& 0.4& 2\\
\end{bmatrix}$, respectively.\\
\\
The return of the risky assets under Regime 1 follows a multi-variate normal distribution $\mathcal{N}([0.01, 0.005, 0.005]^T, 10^{-5}*\begin{bmatrix}
5& 1& 1\\
1& 3& 1\\
1& 1& 3\\
\end{bmatrix})$, and that under Regime 2 follows 
$\mathcal{N}([-0.005, 0.00, -0.01]^T, 10^{-5}*\begin{bmatrix}
8& 1& 1\\
1& 5& 1\\
1& 1& 5\\
\end{bmatrix}$). \\
\\
The initial prices of the three assets are \$3, \$2, and \$3, respectively. The regime transition matrix is $\begin{bmatrix}
0.95& 0.05\\
0.08& 0.92\\
\end{bmatrix}$, with an average of 61.5\% of the time under Regime 1 and 38.5\% of the time under Regime 2. Goal of the problem is to maximize the CRRA utility of terminal wealth with risk aversion parameter $\gamma=-1$.\\
\\
Given the parameters, we create the approximated orthogonal portfolios based on the permanent transaction cost matrices, as they are a better description of market impact than the temporary costs. 
When large amounts are traded, the market impact is likely to be a convex function, and therefore we need an "average" permanent transaction cost matrix for calculating the approximate orthogonal portfolio.\\
\\
On average we need to sell $\frac{20}{10}=2$ chunks of asset 1 in each period, which results in a permanent transaction cost of $61.5\%(10^{-4}*(2)+10^{-4}*(2^2)) + 38.5\%(10^{-4}*2*(2)+10^{-4}*3*(2^2)) = 0.0985\%$ on itself. Similarly calculating other permanent transaction costs, we get the average permanent transaction cost matrix $10^{-4}*\begin{bmatrix}
9.8462& 2.2154& 1.2923\\
1.6615& 12.9231& 1.2923\\
1.2923& 1.2923& 8.3077\\
\end{bmatrix}$. Eigen decomposition suggests the approximate orthogonal portfolios to be 
\begin{itemize}
    \item P\#1 = 0.488 units asset\#1 + 0.826 units asset\#2 + 0.281 units asset\#3
    \item P\#2 = 0.765 units asset\#1 - 0.484 units asset\#2 + 0.425 units asset\#3
    \item P\#3 = -0.429 units asset\#1 - 0.084 units asset\#2 + 0.899 units asset\#3.
\end{itemize}
To accomplish the goal of selling 20 chunks of each of the original assets, we need to liquidize 31.13 chunks of portfolio P\#1, 10.52 chunks of P\#2 and 7.53 chunks of P\#3.\\
\\
We optimize the trading schedule of portfolios P\#1, P\#2 and P\#3 using the dynamic program algorithm for single security. When dynamic program tells us to sell $p_1, p_2$ and $p_3$ chunks of these approximate orthogonal portfolios, respectively, we immediately learn that we shall trade $[p_1,p_2,p_3]^T
\begin{bmatrix}
0.488 & 0.826 & 0.281\\
0.765 & -0.484 & 0.425 \\
-0.429 & -0.084 & 0.899
\end{bmatrix}^{-1}$ of the original assets, respectively.

\subsubsection{Results with orthogonal portfolios alone}
\begin{figure}[ht!]
\center
\begin{subfigure}{0.48\textwidth}
\includegraphics[width=0.9\linewidth]{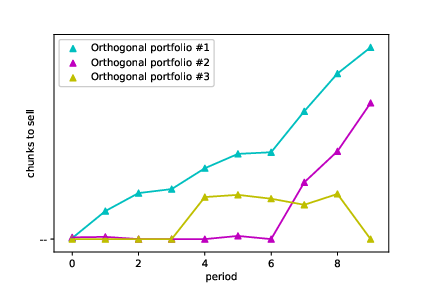}
\caption{The chunks of orthogonal portfolios to sell in each period.}
\end{subfigure}
\begin{subfigure}{0.48\textwidth}
\includegraphics[width=0.9\linewidth]{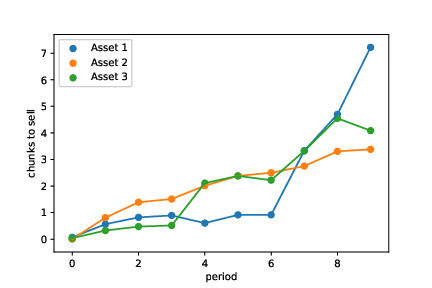} 
\caption{The number of chunks of original assets to sell in each period.}
\end{subfigure}
\caption{Trading schedule in terms of approximate orthogonal portfolios and in terms of original assets of interest, when the whole horizon is under Regime 1.}\label{fig:3asset_regime1}
\end{figure}

We are able to achieve average terminal wealth of \$159.919 based on 10000 simulations, with expected utility -0.00626. Below we illustrate a few representative realized paths. The amounts to sell depend on current underlying regime, accumulated cash so far as well as the current price of the risky assets. Yet we notice that the underlying regime has the most significant influence on the selling schedule among all realized paths. Therefore, the realized paths are described in terms of realized regimes. \\
\\
\textbf{Example a.1: Entire period under Regime 1.}
In the case that the entire period is under Regime 1, the returns of the risky assets are relatively high, and the transaction costs of both types are relatively low compared to Regime 2. There is benefit of holding the assets for longer. A sample realized selling schedule appears in Figure \ref{fig:3asset_regime1}.

\begin{figure}[ht!]
\center
\begin{subfigure}{0.48\textwidth}
\includegraphics[width=0.9\linewidth]{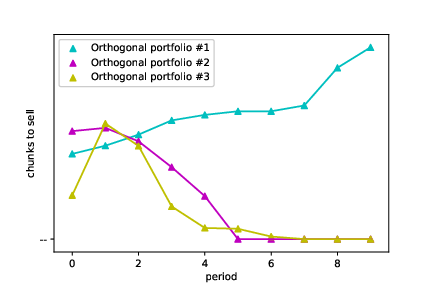}
\caption{The chunks of orthogonal portfolios to sell in each period.}
\end{subfigure}
\begin{subfigure}{0.48\textwidth}
\includegraphics[width=0.9\linewidth]{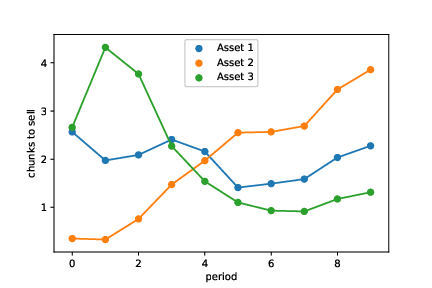} 
\caption{The number of chunks of original assets to sell in each period.}
\end{subfigure}
\caption{Trading schedule in terms of approximate orthogonal portfolios and in terms of original assets of interest, when the whole horizon is under Regime 2.}
\label{fig:3asset_regime2}
\end{figure}

\begin{figure}[ht!]
\center
\begin{subfigure}{0.48\textwidth}
\includegraphics[width=0.9\linewidth]{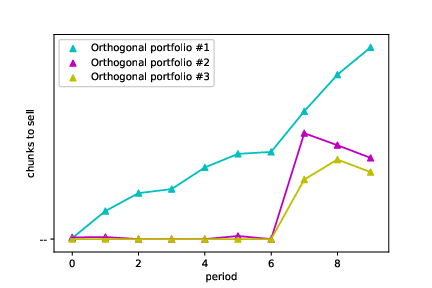}
\caption{The chunks of orthogonal portfolios to sell in each period.}
\end{subfigure}
\begin{subfigure}{0.48\textwidth}
\includegraphics[width=0.9\linewidth]{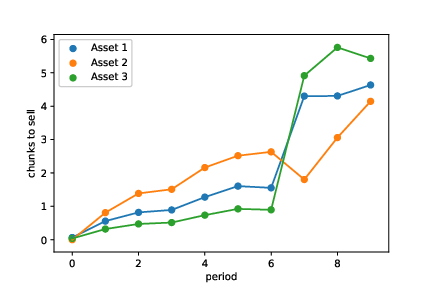} 
\caption{The number of chunks of original assets to sell in each period.}
\end{subfigure}
\caption{Trading schedule in terms of approximate orthogonal portfolios and in terms of original assets of interest, when the first 6 periods are under Regime 1 and the rest under Regime 2.}
\label{fig:3asset_mixed}
\end{figure}

\textbf{Example a.2: Entire period under Regime 2.}
When the entire period is under Regime 2, the returns are low and the transaction costs increases rapidly with number of chunks sold. Hence, intuitively one should sell the assets in a more even pace over the horizon. A sample realized strategy is included in Figure \ref{fig:3asset_regime2}.
\\
\\
\noindent
\textbf{Example a.3: The first six periods under Regime 1, followed by four periods under Regime 2.}
There is a mixture of regimes. Since our strategy obeys non-anticipativity constraint, we will not look ahead. As shown in Figure \ref{fig:3asset_mixed}, the selling schedule in the first six periods are therefore similar to that in Example a.1 where the whole horizon is under Regime 1. In the last four periods, selling are spread more evenly due to the regime switch.
\\

\subsubsection{Improve with neural networks}
\begin{figure}[ht!]
\center
\includegraphics[width=0.6\linewidth]{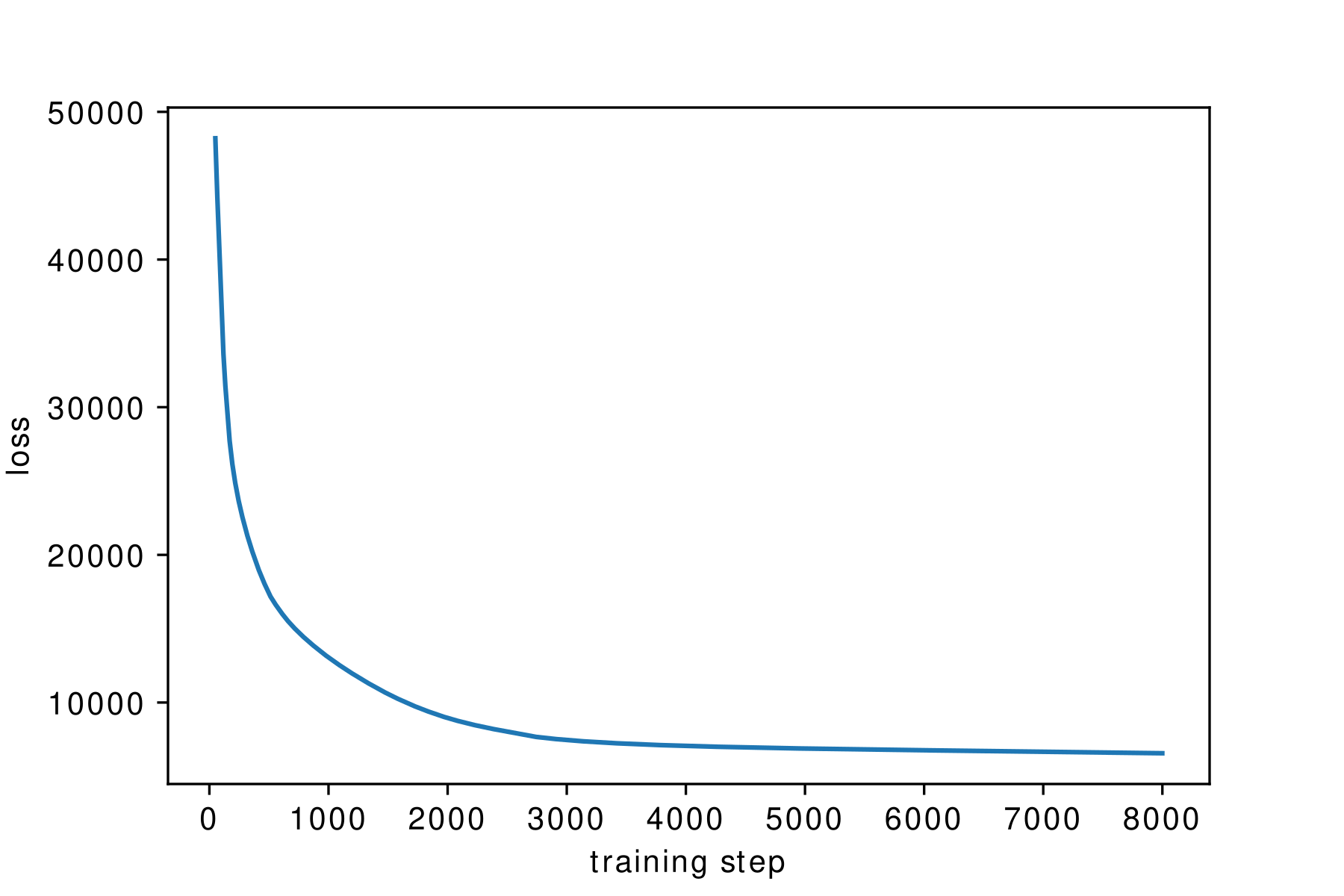} 
\caption{Loss of pre-training.}
\label{loss_pretrain}
\end{figure}

We employ a neural network with four neurons in the hidden layer. First we pretrain the weights with the solution provided by the orthogonal portfolio method. The input features include time period, current regime, amounts left to be sold, current prices and accumulated cash amount. The goal of the pretrain is to minimize the mean squared error between output and the solution from orthogonal portfolio method. Twenty thousand steps are trained using gradient method with learning rate $10^{-7}$, where the loss function appears in Figure \ref{loss_pretrain}. 
\\
\\
\noindent
Then we optimize on the original model. We initilize the neural network with pretrained weights, and train 1,000 steps with Adam Optimizer based on gradient method, resulting in a selling strategy with expected terminal wealth \$160.829 and expected utility -0.00622. The average wealth and utility are shown in Figure \ref{fig:nn}.  Comparing to the average wealth \$159.919 and expected utility -0.00626 from approximated orthogonal portfolios alone, we notice a fair improvement by utilizing the neural networks.
\begin{figure}[ht!]
\center
\begin{subfigure}{0.48\textwidth}
\includegraphics[width=0.9\linewidth]{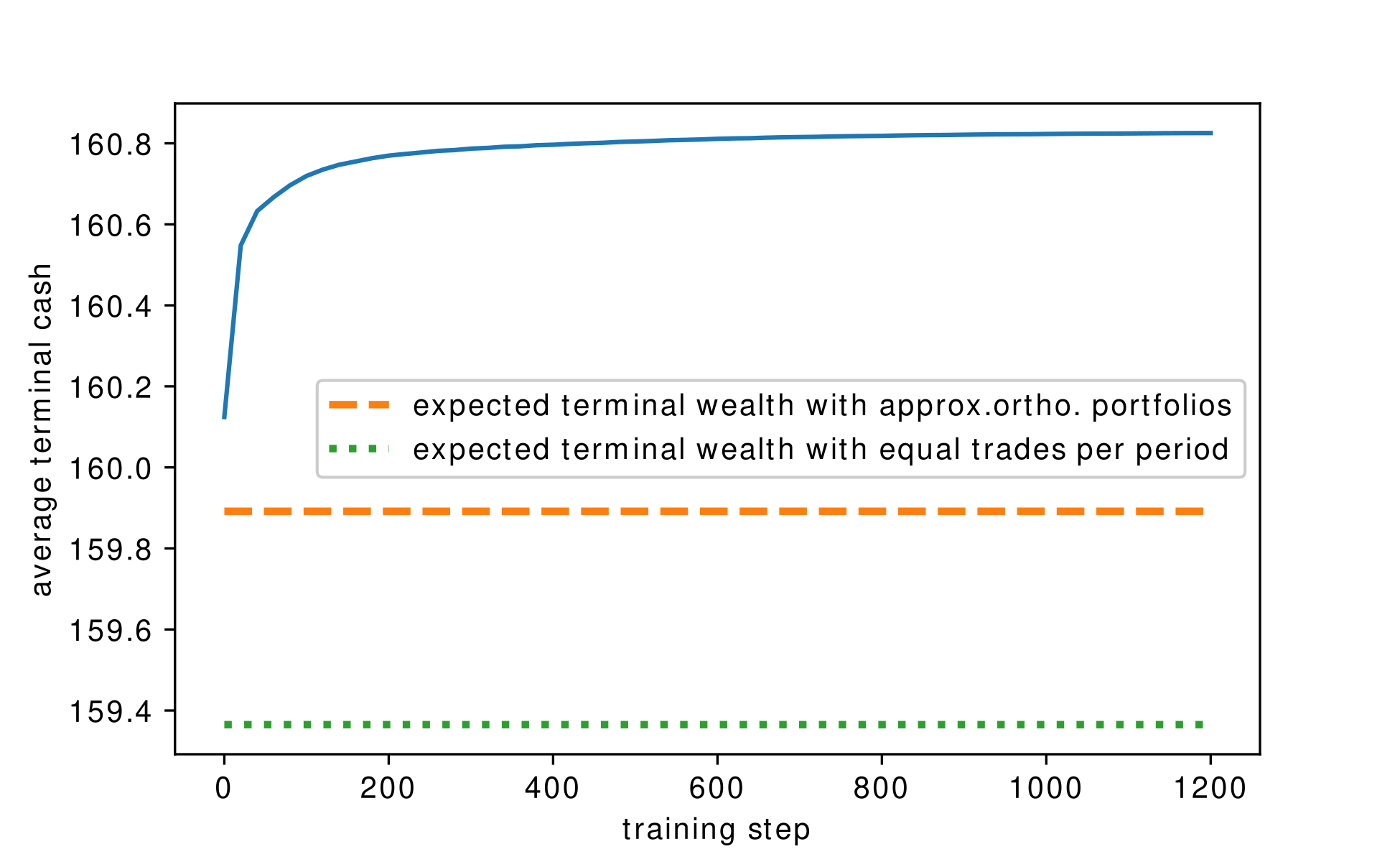}
\caption{Average terminal wealth at training steps.}
\end{subfigure}
\begin{subfigure}{0.48\textwidth}
\includegraphics[width=0.9\linewidth]{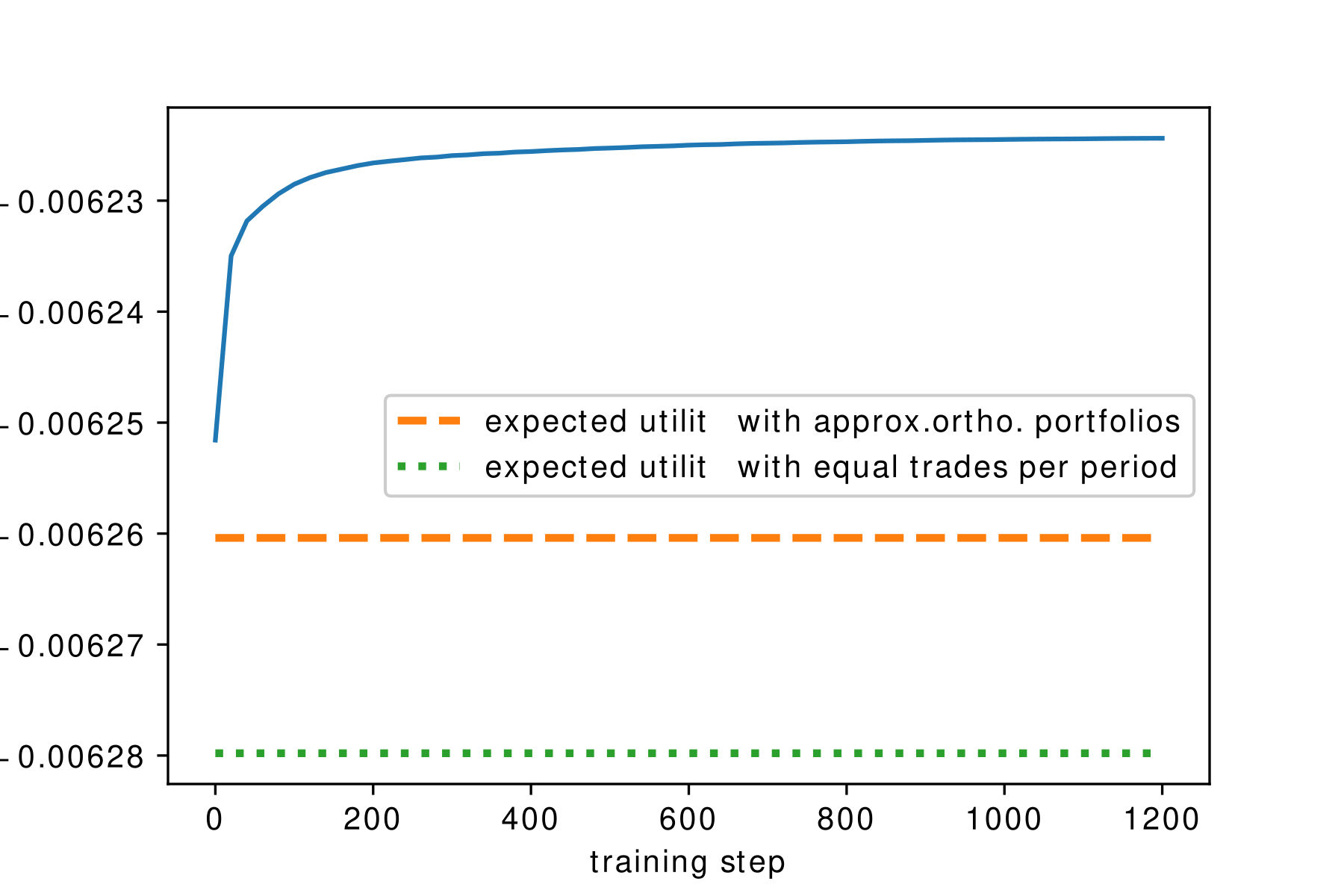} 
\caption{Average terminal utility at training steps.}
\end{subfigure}
\caption{Training of neural network.}
\label{fig:nn}
\end{figure}

\subsection{An example with 4 regimes}
\begin{figure}[ht!]
    \centering
    \includegraphics[width=0.6\textwidth]{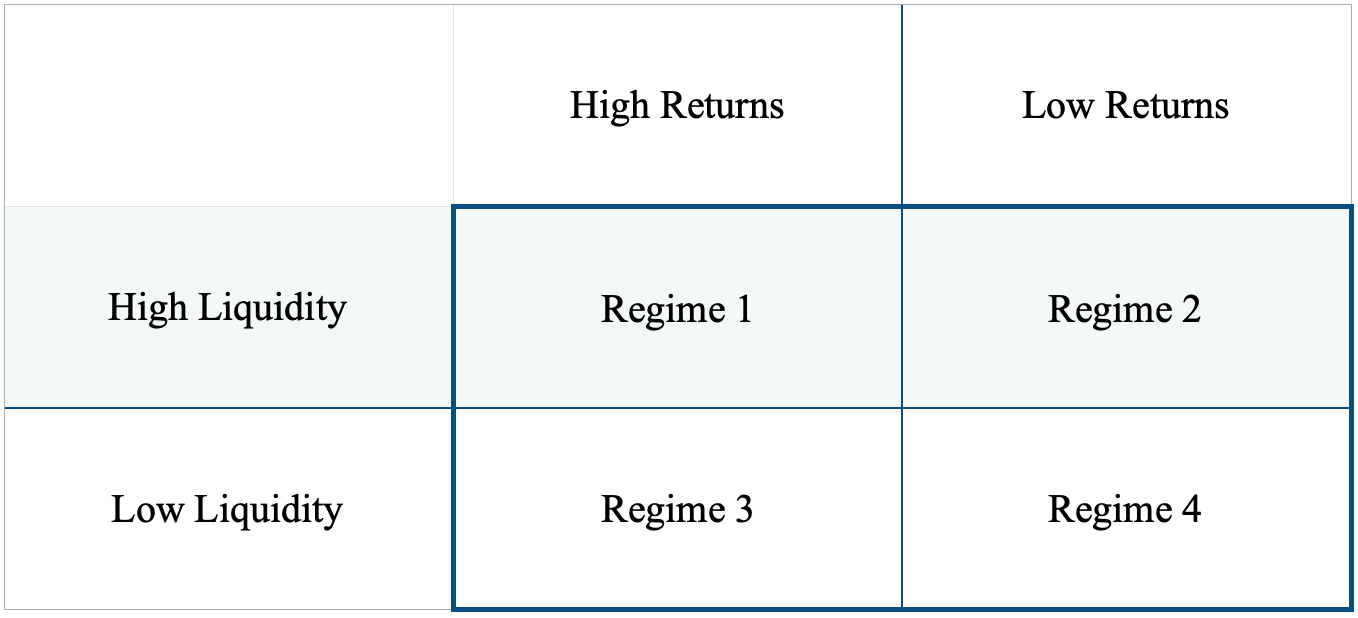}
    \caption{Qualitative description of the four regimes.}
    \label{4regime}
\end{figure}

Increasing the number of regimes deteriorates the accuracy of estimated trading costs with approximated orthogonal portfolios, as more deviations from the average level are expected. The suggested strategy by approximated orthogonal portfolios are therefore expected to be relatively suboptimal. In this subsection, we extend the previous 3-asset example to a four-regime case. In particular, the temporary transaction costs, permanent transaction costs, as well as return dynamics are kept the same under Regime 1 and Regime 2. Under Regime 3, the returns follow the same distribution as under Regime 1, but all transaction costs are doubled. Under Regime 4, the returns follow the same distribution as under Regime 2, but all transaction costs are doubled. The regimes can be interpreted qualitatively as Figure \ref{4regime}. The transition matrix is assumed to be $\begin{bmatrix}
0.8& 0.15& 0.05& 0\\
0.2& 0.75& 0& 0.05\\
0.08& 0& 0.8& 0.12\\
0& 0.08& 0.32& 0.6\\
\end{bmatrix}$.\\

\begin{figure}[ht!]
\center
\begin{subfigure}{0.48\textwidth}
\includegraphics[width=0.9\linewidth]{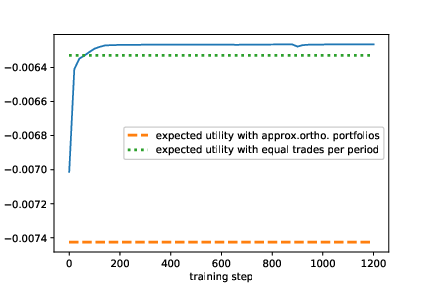}
\caption{Average terminal wealth at training steps.}
\end{subfigure}
\begin{subfigure}{0.48\textwidth}
\includegraphics[width=0.9\linewidth]{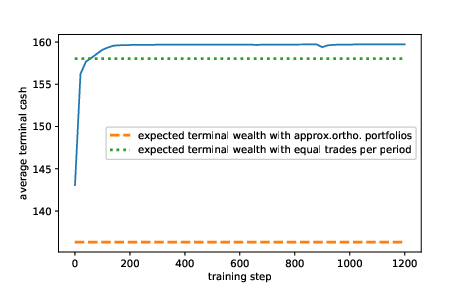} 
\caption{Average terminal utility at training steps.}
\end{subfigure}
\caption{Training of neural network.}
\label{fig:nn_4regime}
\end{figure}

\noindent
We calculate the average permanent transaction costs, from which approximate orthogonal orthogonal portfolios are derived. Each portfolio is treated as a single asset and the optimal trading schedule is calculated. Combining results from all three orthogonal portfolios, we achieve an average wealth of \$136.319 and average utility of -0.007425. Note that before the selling starts, the 3 assets worth a total of \$160, much more than we gain with approximated orthogonal portfolios alone. As a second stage, we will feed this solution to the neural network, and correct the errors that have been made in the estimation process.\\
\\
We train the neural network for 1200 steps, resulting in an average terminal wealth \$159.639 with expected utility -0.006265 (Figure \ref{fig:nn_4regime}). The heap of utility from approximated orthogonal portfolios justifies its goodness as a starting point. On the other hand, though the benchmark of equal trade per period provides higher expected utility than approximated orthogonal portfolios alone, its power of serving as a starting point is deteriorated by the loss of understanding of asset dynamics.\\

\subsection{An example with 10 assets}
\begin{figure}[ht!]
\center
\begin{subfigure}{0.48\textwidth}
\includegraphics[width=0.9\linewidth]{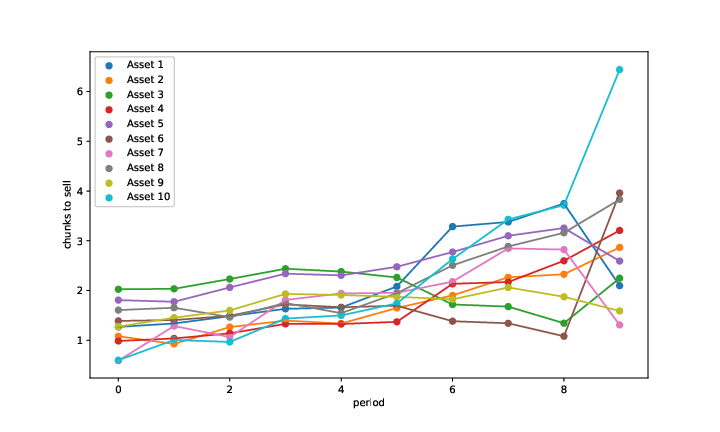}
\caption{A sample selling schedule where the whole horizon are realized in Regime 1.}
\end{subfigure}
\begin{subfigure}{0.48\textwidth}
\includegraphics[width=0.9\linewidth]{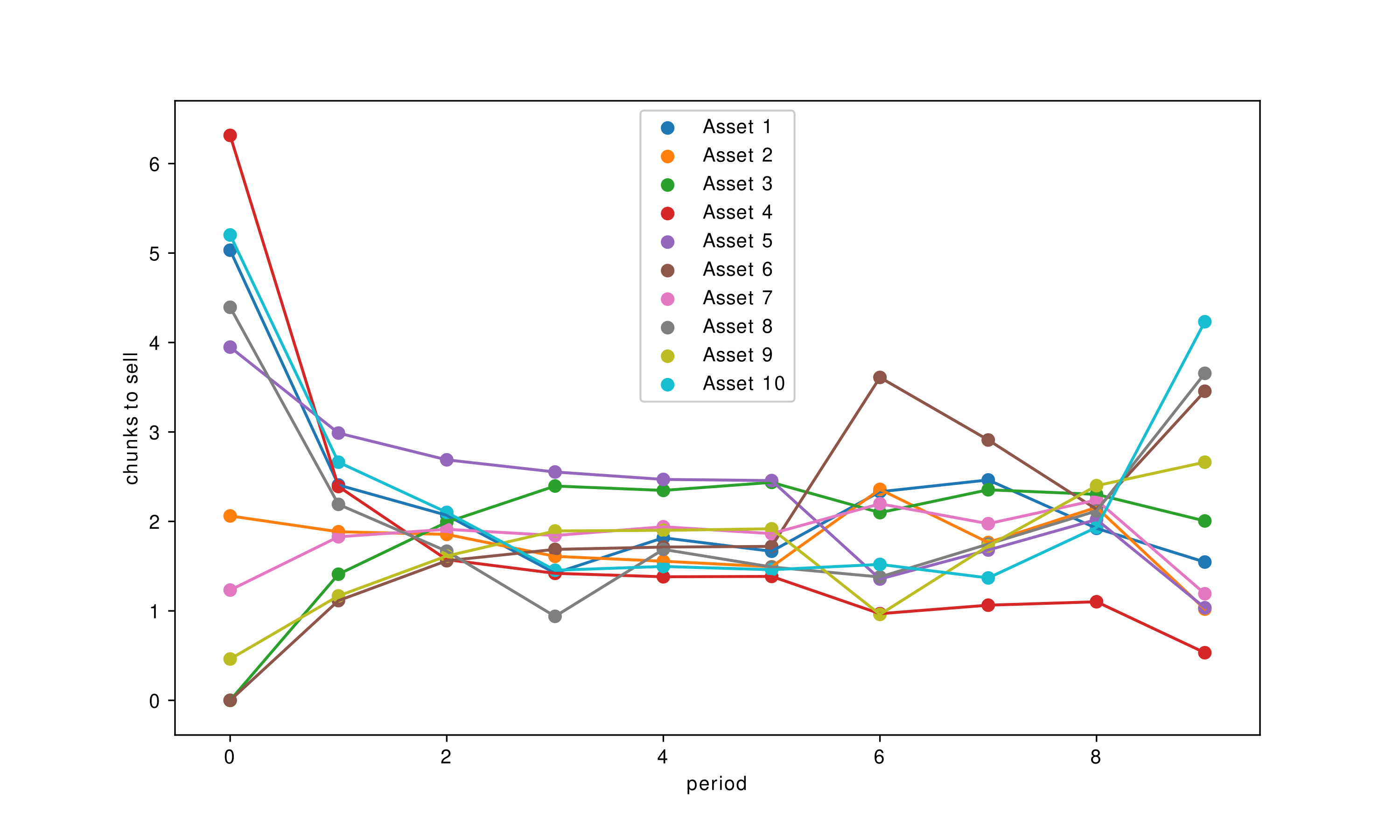} 
\caption{A sample selling schedule where the whole horizon are realized in Regime 2.}
\end{subfigure}
\caption{Illustrations of selling strategy for the 10 asset example. $\gamma=-1$.}
\label{10assets}
\end{figure}

To see how the method performs on more assets, we implement the procedure to an example with 10 assets involved. Twenty chunks are to be sold for each of the assets in ten periods. The matrices for temporary and permanent transaction costs have the same form as before, and are generated with random simulation. The average returns and covariance matrices under two regimes are also generated randomly, with Regime 1 enjoys relatively higher average returns and lower covariances. The parameters are included in the appendix. The initial wealth of the portfolio is \$638.22.\\
\\
We simulate the performance on 10,000 sample paths, whose regimes of the first period follow the stationary distribution of the regime Markov chain. When $\gamma=-1$, with approximated orthogonal portfolios alone, an agent earns \$637.41 on average after 10 periods, with an expected utility -0.001605. As Figure \ref{10assets} suggests, when current under Regime 1, the agent tends to benefit from holding the assets and selling in the future; whereas under Regime 2, the selling schedule is more spread out throughout the horizon.\\

\begin{figure}[ht!]
\center
\begin{subfigure}{0.48\textwidth}
\includegraphics[width=0.9\linewidth]{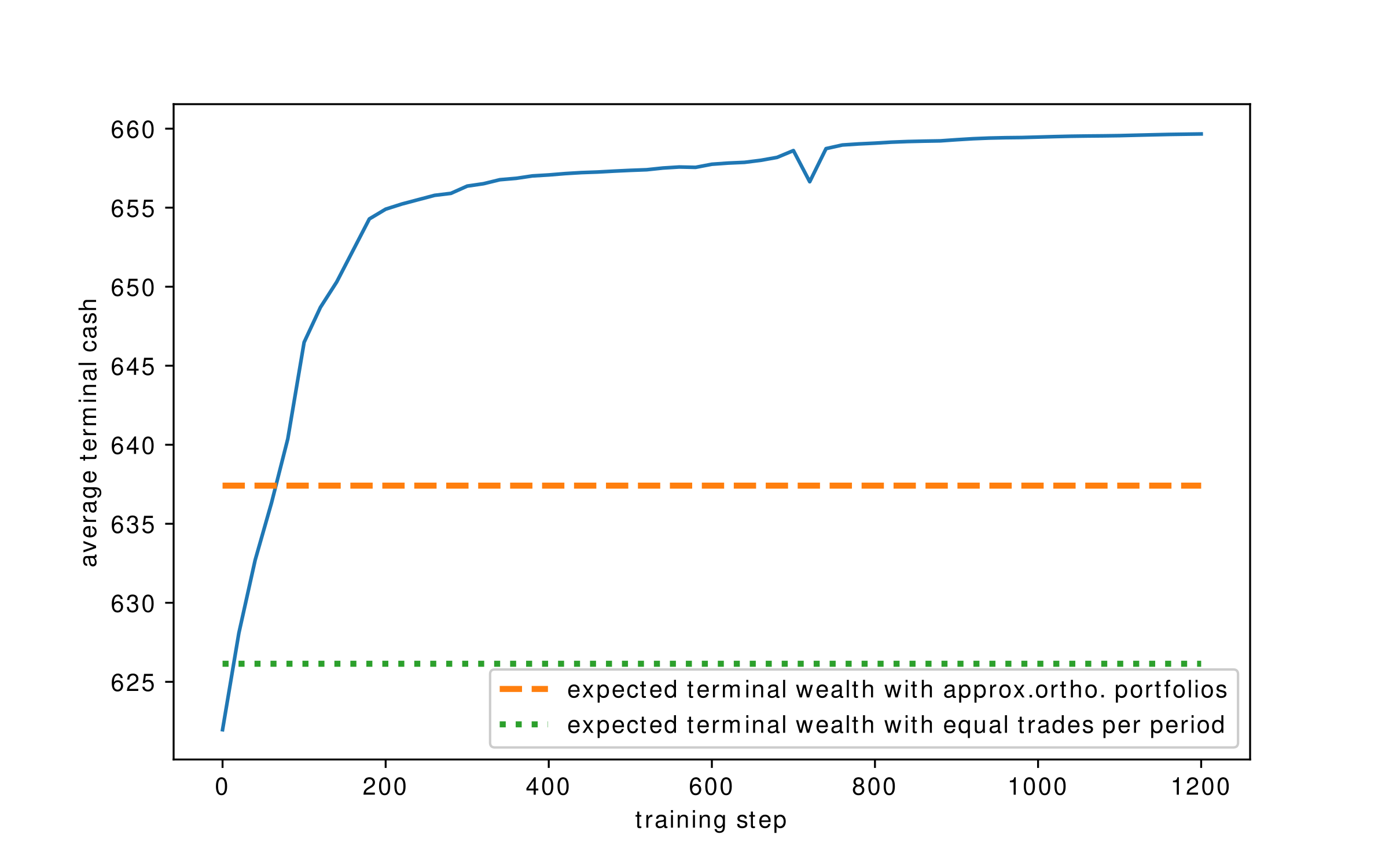}
\caption{Average terminal wealth at training steps.}
\end{subfigure}
\begin{subfigure}{0.48\textwidth}
\includegraphics[width=0.9\linewidth]{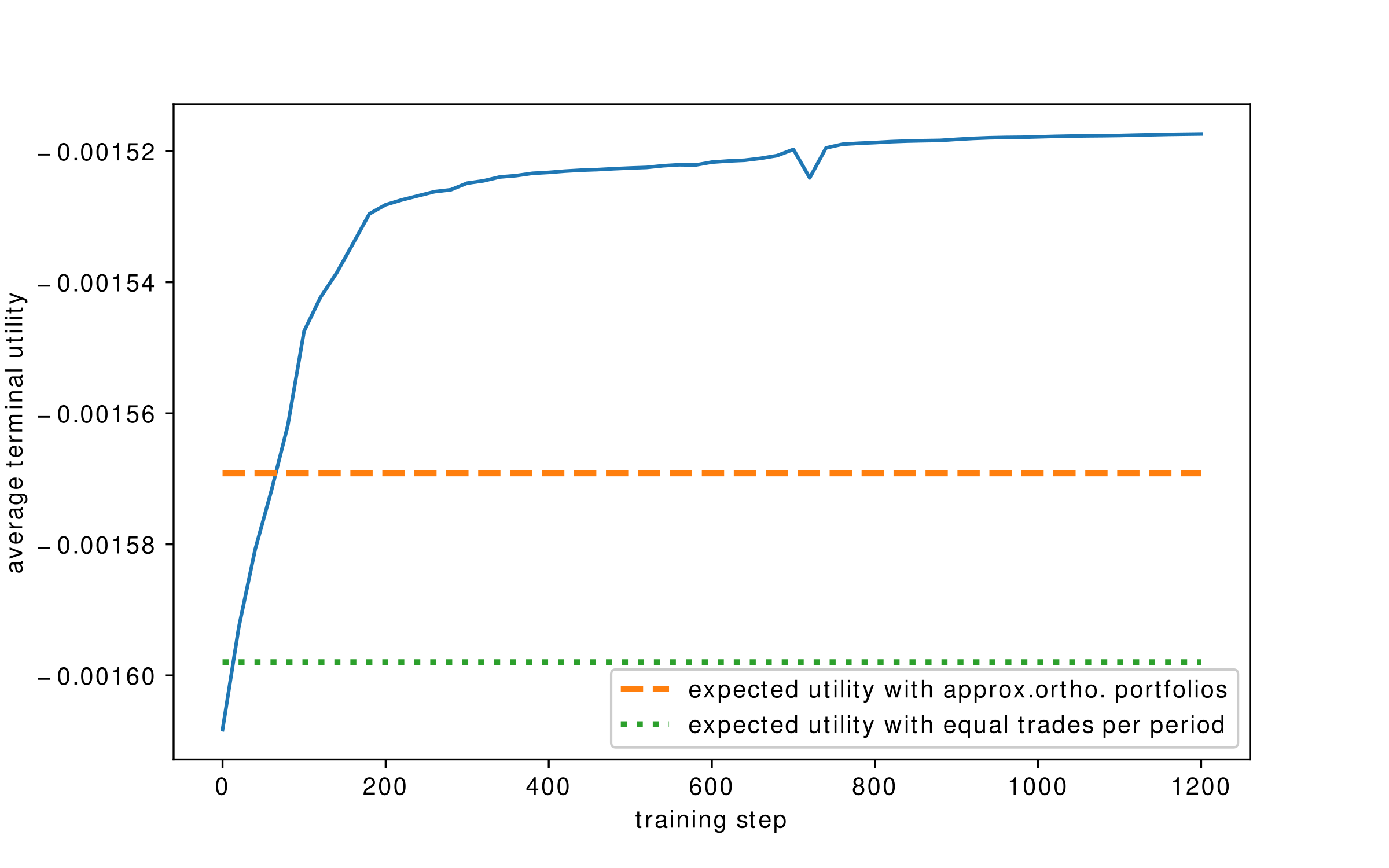} 
\caption{Average terminal utility at training steps.}
\end{subfigure}
\caption{Training of neural network. $\gamma=-1$.}
\label{fig:nn_10assets}
\end{figure}

\noindent
Then we pre-train the weights of a neural network with one hidden layer that contains 7 neurons, aiming to match the strategy suggested by the approximated orthogonal portfolio. The neural network is then trained to optimize on the utility based on the original problem. After 1200 training steps, we achieved an average terminal wealth of \$659.705, with expected utility 0.001517 (Figure \ref{fig:nn_10assets}), which is about a 5.5\% improvement from the strategy learned by approximate orthogonal portfolios alone.\\
\\

\begin{figure}[ht!]
\center
\includegraphics[width=0.8\textwidth]{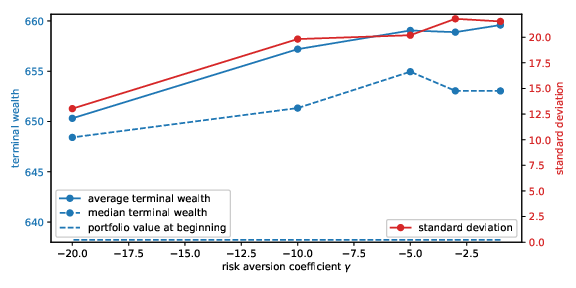}
\caption{Average terminal wealth, median terminal wealth and standard deviation of terminal wealth with various risk aversion coefficient $\gamma = -1, -3, -5, -10$ and $-20$.}
\label{fig:10asset_gamma}
\end{figure}

\noindent
With the 10-asset example, we run the combined algorithm with various risk aversion coefficients $\gamma = -1, -3, -5, -10$ and $-20$ (Figure \ref{fig:10asset_gamma}). The less risk averse, the higher expected terminal wealth can be achieved, which is usually associated with higher median terminal wealth as well as volatility. Due to the random seeds, we do not always observe monotonicity in the risk-aversion coefficient, yet the trend is clear. We compare the performance of these strategies to the benchmark strategy, that sells equal amount at each period (Table \ref{tab:10asset_gamma}). We find that strategies learned with CRRA utility enjoy significantly higher returns, yet suffer from volatility. We note that with a second order Tayler expansion, the expected CRRA utility can be approximated with $\mathbb{E}[U(W)] = U(\mathbb{E}[W]) + \frac{1}{2}U''(\mathbb{E}[W])Var(W)$. Since $U''(\mathbb{E}[W]) = (\gamma-1)\mathbb{E}[W]^{\gamma-2}$, when the total value of the portfolio is high, the impact of variance term in the optimization problem vanishes. Such a strategy is suitable for the traders with low risk aversion and high return expectations. In the next section, we address the trade-off between return and uncertainty.

\begin{table}[ht!]
    \centering
    \begin{tabular}{lp{2.5cm}p{2cm}p{2cm}p{2cm}}
    \toprule
    {}&  Average terminal wealth & Median terminal wealth & Standard deviation\\
    \hline
    CRRA $\gamma=-1$ & 659.525 & 652.873 & 21.492\\
    CRRA $\gamma=-3$ & 659.036 & 653.574 & 21.836\\
    CRRA $\gamma=-5$ & 658.958 & 654.669 & 20.167\\
    CRRA $\gamma=-10$ & 657.195 & 651.364 & 19.844\\
    CRRA $\gamma=-20$ & 650.257 & 648.400 & 12.986\\
    Benchmark: Equal trade each period & 626.143 & 629.729 & 11.087\\
    \bottomrule
    \end{tabular}
    \caption{Average terminal wealth, median terminal wealth and standard deviation of terminal wealth with various risk aversion coefficient $\gamma = -1, -3, -5, -10$ and $-20$.}
    \label{tab:10asset_gamma}
\end{table}

\section{Mean-Variance Optimization}
In the previous section, we scrutinize the performance of our combined method on CRRA utility functions. To address the traders who look for a better balance between expected gain and volatility, we introduce the mean-variance optimization 
\begin{align}\label{mvo}
     Maximize\hspace{0.5cm} \mathbb{E}[W] - \lambda Var(W), 
\end{align}
where $W$ is the terminal wealth, and $\lambda$ is the mean-variance risk aversion coefficient.\\
\\
\subsection{Results with various mean-variance coefficients}
\begin{figure}[ht!]
\center
\includegraphics[width=0.8\textwidth]{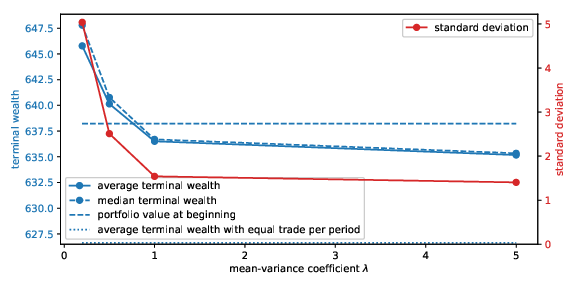}
\caption{Average terminal wealth, median terminal wealth and standard deviation of terminal wealth with various mean-variance coefficient $\lambda$.}
\label{fig:10asset_mvo}
\end{figure}

We continue with the 10-asset example, and optimize problem (\ref{mvo}) with $\lambda = 0.2, 0.5, 1$ and $5$ (Figure \ref{fig:10asset_mvo}). For the starting strategy, we still use the solution learned with the orthogonal portfolios. It turns out to provide a superior starting point than naively trading equal amount per period. We train a neural network with the same structure as previously for each $\lambda$. The metrics including mean and standard deviation appear in Table \ref{tab:10asset_mvo}. With mean-variance optimization, we provide a better balancing strategy between expected terminal wealth and risk control. In particular, the mean-variance optimal trading schedules achieve higher expected wealth than the benchmark strategy that sells equal amount per period, with significantly lower standard deviation compared to the benchmark.

\begin{table}[ht!]
    \centering
    \begin{tabular}{lp{2.5cm}p{2cm}p{2cm}p{2cm}}
    \toprule
    {}&  Average terminal wealth & Median terminal wealth & Standard deviation\\
    \hline
    Mean-variance $\lambda=0.2$ & 645.784 & 647.796 & 5.038\\
    Mean-variance $\lambda=0.5$ & 640.141 & 640.771 & 2.510\\
    Mean-variance $\lambda=1$ & 636.505 & 636.690 & 1.540\\
    Mean-variance $\lambda=5$ & 635.172 & 635.338 & 1.404\\
    Benchmark: Equal trade each period & 626.143 & 629.729 & 11.087\\
    \bottomrule
    \end{tabular}
    \caption{Average terminal wealth, median terminal wealth and standard deviation of terminal wealth with various mean-variance coefficient $\lambda$.}
    \label{tab:10asset_mvo}
\end{table}

\subsection{Mean-variance efficient frontier}
\begin{figure}[ht!]
\center
\includegraphics[width=0.8\textwidth]{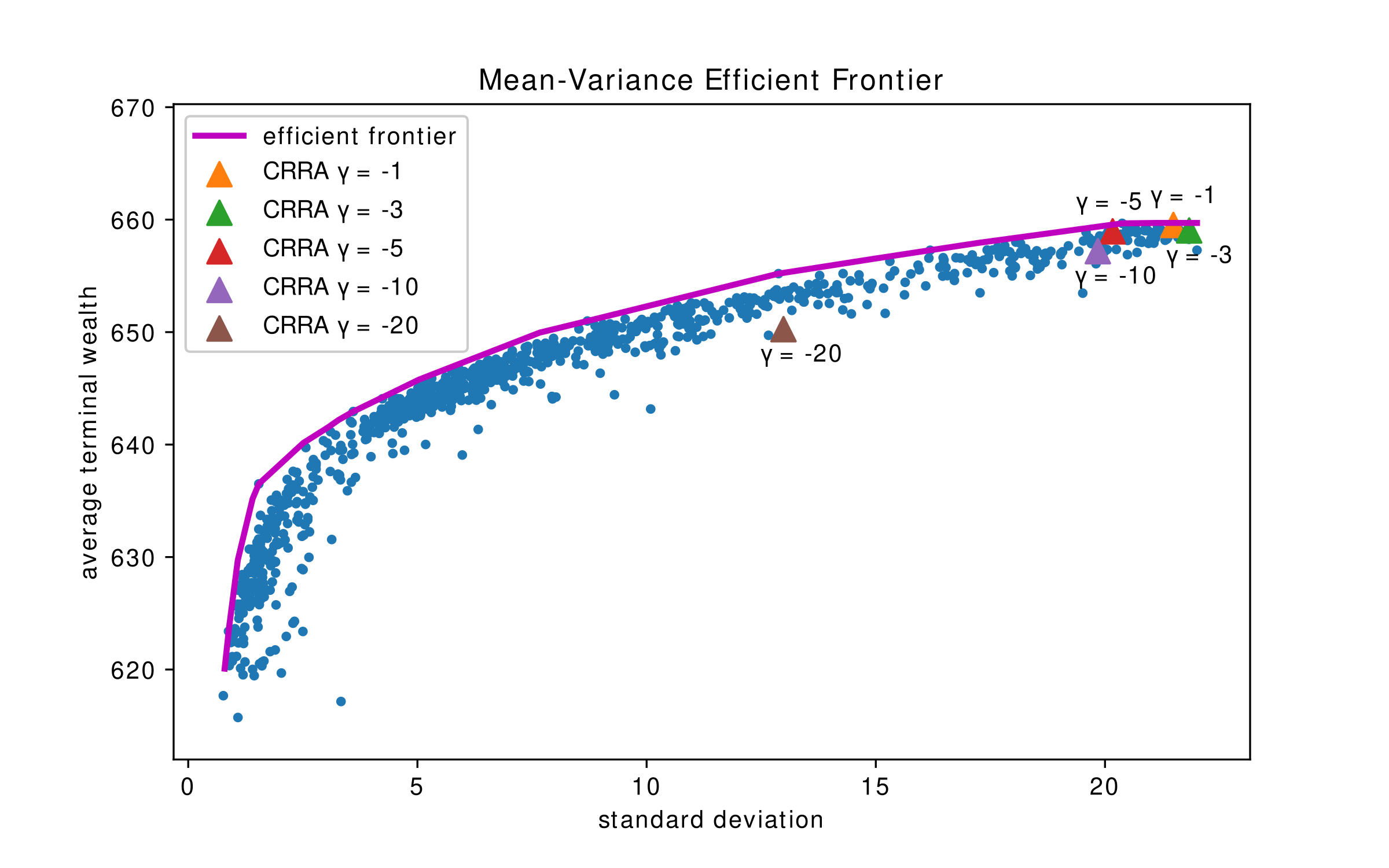}
\caption{Mean-variance efficient frontier.}
\label{efficientfrontier}
\end{figure}

In addition to the representative mean-variance coefficients, we apply the combined algorithm to a wide range of $\lambda$'s from 0 to 10 in order to plot the mean-variance efficient frontier. Since the random seeds in the neural network brings different solutions every time, we train the neural network for 20 times for each $\lambda$ to attain the best efficient frontier. All the results are recorded in Figure \ref{efficientfrontier}. The solutions that are inside the efficient frontier, although suboptimal, are all close to the frontier and provide near optimal performance, justifying the stability of neural network performance.\\
\\
CRRA solutions are not mean-variance optimal, as shown in the figure. This is because CRRA utility counts upside and downside volatility asymmetrically, whereas mean-variance optimization treats both sides the same. We provide both formulations in the paper to advocate the power of the combined method in optimal execution of a portfolio. Traders may choose either objective function that better fits their risk preferences.\\

\subsection{Extension to more periods}\label{20period}
\begin{figure}[ht!]
\center
\includegraphics[width=0.8\textwidth]{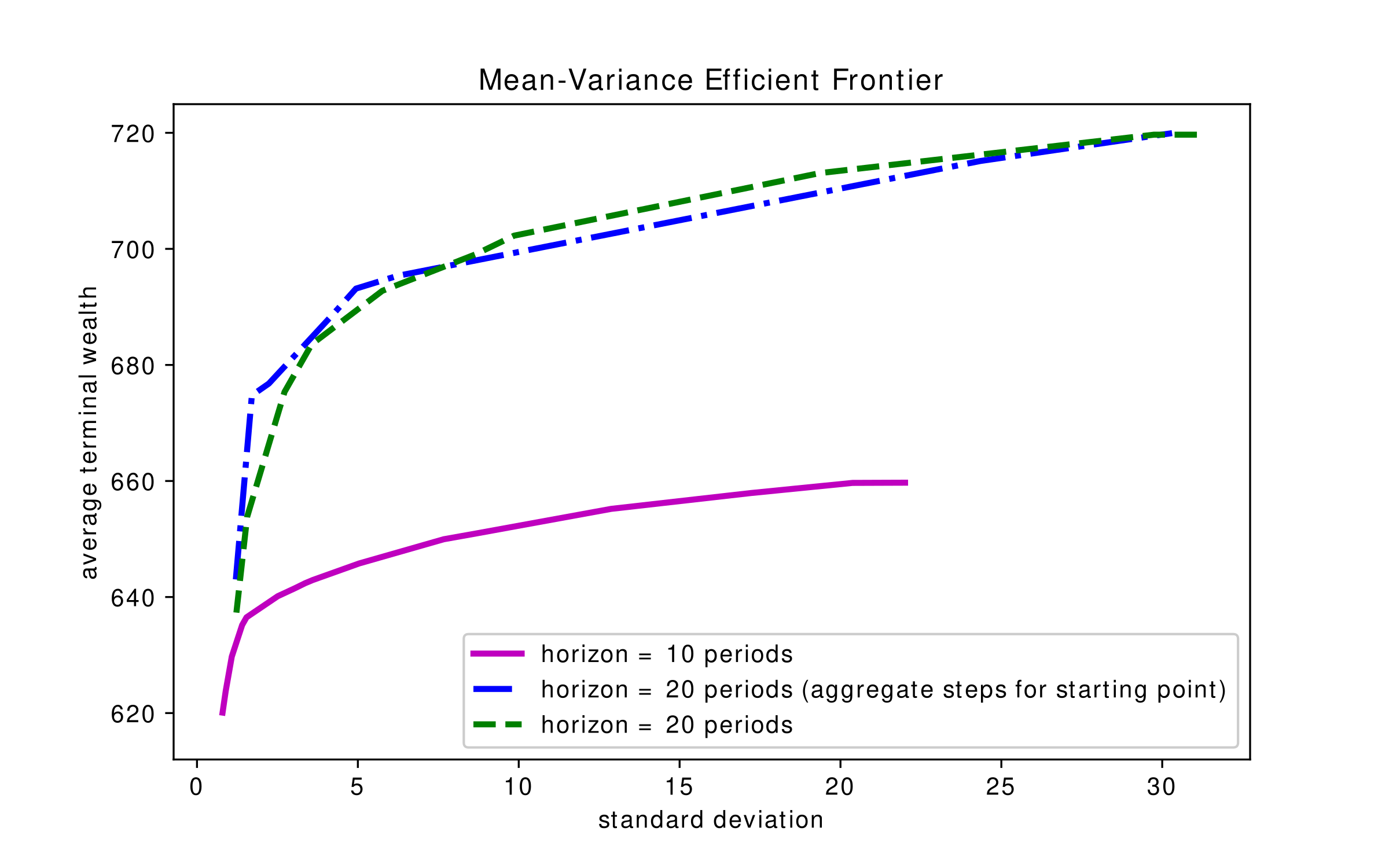}
\caption{Mean-variance efficient frontier.}
\label{efficientfrontiercomparison}
\end{figure}

Now we extend the 10-asset model to allow the trade selling the portfolio in 20 periods. By allowing a longer planning window, the feasible set of viable trading strategies gets larger, and therefore we can expect the efficient frontier to move top-left. \\
\\
With consideration of running time, we can improve the efficiency by aggregating steps when learning the selling schedules of approximated orthogonal portfolios. For the 20-period example, we aggregate the time steps in groups of two, i.e., periods $\{1,2\},\{3,4\},...$ and $\{19,20\}$, and assume that the trader sells equal amount of assets at all steps in the same group. Since our dynamic program part has running time proportional to the size of the state space provided the fixed size of the action space, we expect the grouping technique to reduce the running time of dynamic program part to decrease by nearly a half. The details can be found in Section \ref{runningtime}.\\
\\
We create the efficient frontier with running (i) the full 20-period model and (ii) the grouped 10-period model on approximated orthogonal portfolios as the starting point for neural networks. With the solutions from approximated orthogonal portfolios alone, (i) provides an average terminal wealth of \$667.463 where as (ii) ends up with \$626.631 in the case CRRA risk aversion $\gamma=-1$. There is loss from aggregating steps in the starting solution. However, the trained results of neural networks are similar in both cases, and therefore we conclude that it is effective to aggregate steps when the trading horizon is long. However, we also note that the extreme case does not provide much advantage as a starting point, where the trader groups all periods into one and assuming equal trades per period. Thus, while we recommend aggregating steps when selling horizon is long, one should be cautious not to over-simplify the model. The comparison of efficient frontiers we find by the combined method appear in Figure \ref{efficientfrontiercomparison}. It is consistent with the claim that having a longer selling horizon leads to a better efficient frontier. In particular, the maximum achievable expected terminal wealth is raised from \$659.705 in the 10-period model to \$719.948 in the 20-period model.\\

\section{Running time}\label{runningtime}
\subsection{Running time breakdown}
So far, we have introduced an offline algorithm for the optimal execution of a portfolio. By offline learning, we refer to the learning of trading strategy from scratch. In other words, no prior knowledge of optimal execution is provided in the offline case. In order to learn the strategy, our combined strategy suggests a trader to (i) calculate the approximated orthogonal portfolios, (ii) determine the selling schedule of each orthogonal portfolio, (iii) pretrain the neural network weights, and (iv) train the neural network with the original model. In this procedure, there are three parts that are time-consuming: (ii), (iii) and (iv). The time it takes to calculate approximated orthogonal portfolios is negligible, and we will focus our discussion of running time on steps (ii)-(iv).\\
\\
We employ Princeton University TigerCPU HPE Linux Cluster for implementation, and all tasks are completed with a single node. The specification of each step is as follows:
\begin{itemize}
    \item Step (ii): It is the most time-consuming part of the combined algorithm. We run all models with three iterations where each simulates 1,000 paths at all states. Note that the strategy of one orthogonal portfolio is independent of that of other portfolios. Therefore, if one allows parallel running algorithm, she learns the trading schedule of all approximated orthogonal portfolios in $O(1)$. When parallel running is disallowed, running time of step (ii) is linear in number of underlying assets.
    \item Step (iii): Pretraining of the weights usually converge within 8,000 steps.
    \item Step (iv): Training the neural network on the original model usually converges within 1,200 steps.
\end{itemize}

The running time breakdown on the 10-asset model with different hyperparameters appears in Table \ref{tab:runningtime}. The learning time on each approximated orthogonal portfolio is proportional to the number of periods. Neural network training time grows at a close to linear rate. Therefore, when there is no step aggregation, our model takes approximately linear running time in the length of trading period. On the other hand, as our example in Section \ref{20period} suggests, by aggregating steps carefully, the running time can be significantly decreased and still provides promising result.\\
\begin{table}[ht!]
\small
    \centering
    \begin{tabular}{l|p{3.6cm}p{3.6cm}p{3.6cm}}
    \toprule
    Step  & 10-period & 20-period \newline(full model) & 20-period \newline(aggregated into 10-steps for Step (ii))\\
    \hline
    \hline
    (ii) & 312 sec; or \newline 31.2 sec per ortho.port. & 669 sec; or \newline 66.9 sec per ortho.port. & 316 sec; or \newline 31.6 sec per ortho.port.\\
    \hline
    (iii): 8000 steps & 10.3 sec & 23 sec & 23 sec\\
    \hline
    (iv): 1200 steps & 16.4 sec & 35.9 sec & 35.9 sec \\
    \bottomrule
    \end{tabular}
    \caption{Running time of each step on the 10-asset model with various planning horizon and dynamic program size.}
    \label{tab:runningtime}
\end{table}

\noindent
In the above analysis, we do not employ any early stopping rules. One could further shorten the running time by applying early stopping rule to terminate the training once it converges.

\subsection{Efficiency improvements}
For institutional traders who face large volumes of portfolios to be implemented every day, an offline learning algorithm could be infeasible 
even when each trading strategy takes only about one minute. Here, we discuss a few possibilities of improving the computational efficiency of our algorithm. We will not explicitly implement these improvement methods, as they require large amount of experiments. However, they provide a hint of how institutional traders may employ the combined method with most efficiency.\\
\\
\begin{itemize}
    \item \textbf{Improvement method 1}: shorten the running time on Step (ii) if parallel running is not available.\\ 
    When parallel computing is not available, the running time of Step (ii) grows linearly in the number of  assets in the original portfolio. Here, we propose a method to reduce the computation.\\
    When calculating the approximated orthogonal portfolios, we also have the information of how many units of each portfolios need to be sold to replicate the original portfolio. Some approximated orthogonal portfolios may have a larger impact than others. For example, in Section \ref{example:3asset}, the original portfolio is equivalent to 31.13 chunks of P\#1, 10.52 chunks of P\#2 and 7.53 chunks of P\#3. If the computing power is limited, we can run Step (ii) only on the first few important approximated portfolios, and assume equal trades per period for the rest.\\
    By fixing the number of important approximated orthogonal portfolios, the running time on Step (ii) can be $O(1)$ even when parallel computing is not available. In addition, by sorting their relative importance, we still keep the most information for the training steps.\\
    
    \item \textbf{Improvement method 2}: shorten the running time on Step (ii), the learning of selling schedules on orthogonal portfolios.\\ 
    First, generate a large number of single-asset experiments, and store the trading strategy suggested by the dynamic programs locally. \\
    Then, train a neural network by supervised learning. The inputs are parameters of the single-asset models, and the outputs are the amount of asset to be sold each period. \\
    After the model is trained, the traders no longer needs to run a dynamic program for each of the orthogonal portfolios. Instead, the traders simply input the parameters of the orthogonal portfolio, and the model will output the strategy.\\
    This improvement method saves time by skipping running dynamic programs for each experiment. In addition, it does not require large memory space. After the supervised learning is complete, we only need to store the weights in the trained model, and no longer need to keep the experiments generated in the first step.\\ 
    \item \textbf{Improvement method 3}: shorten the running time on Step (i)-(iii) and directly provide a starting point for (iv).\\
    First, generate a large number of multi-asset experiments and learn the trading strategy for each. The results should be stored locally.\\
    As new execution requests arise, the traders find the stored experiment with the closest parameters, and use the corresponding stored strategy as the starting point for the neural network in Step (iv).\\
    The merit of this improvement method is that it only keeps the essential step in the original algorithm, which directly optimize on the target model. However, this method requires a large memory space to store the parameters of all ``pre-learned'' experiments. When new execution requests are solved, they shall also be stored for future use, and therefore the required space grows.
\end{itemize}

\section{Conclusions and next steps}
In this paper, we take advantage of approximated orthogonal portfolios, and present a combined method of dynamic program and neural network to tackle the problem of optimal execution of a multi-asset portfolio in a regime-switching market. We include numerical examples with both CRRA utility functions and mean-variance optimization. Trading strategies of both family beat the benchmark of equal trade per period, with CRRA solutions offering higher expected terminal wealth, while mean-variance solutions providing better risk control.\\
\\
This framework of utilizing an approximation numerical method and then employ a neural network for correction can be readily extended to more assets, enabling efficient learning of trading strategies that involves large volumes under multiple underlying regimes. The running time of this combined method is expected to be linear in number of assets as well as in number of regimes, and can be even shortened by allowing parallel computing.\\
\\
For next steps, we are interested in extending the combined method to various market environments, for example, a slow-decaying temporary transaction cost that lasts for a few periods before the price fully rebounds.  \\

\bibliographystyle{abbrvnat}
\bibliography{main}

\appendix
\section{Parameters for the 10 asset example}
The first order term of the temporary transaction costs in Regime 1 is $tr_1^{a,\alpha}= 10^{-3}*$
$$
\begin{bmatrix}
14.799 & 3.424 & 1.118 & -4.783 & -1.469 & -5.108 & -6.774 & 0.422 & -2.064 & 2.037 \\3.424 & 6.140 & 1.004 & 1.282 & -6.005 & 0.180 & -4.363 & -0.368 & 3.165 & 1.737 \\1.118 & 1.004 & 20.056 & -1.669 & -4.683 & 0.728 & 1.572 & 1.707 & 5.723 & 1.516 \\-4.783 & 1.282 & -1.669 & 8.225 & -2.857 & 3.677 & 2.663 & -0.315 & 3.898 & 0.256 \\-1.469 & -6.005 & -4.683 & -2.857 & 15.108 & 0.037 & 3.176 & -4.074 & -5.551 & 0.831 \\-5.108 & 0.180 & 0.728 & 3.677 & 0.037 & 3.420 & 3.328 & -0.088 & 4.059 & 0.182 \\-6.774 & -4.363 & 1.572 & 2.663 & 3.176 & 3.328 & 8.113 & 2.452 & 1.257 & -1.529 \\0.422 & -0.368 & 1.707 & -0.315 & -4.074 & -0.088 & 2.452 & 5.196 & 0.465 & -2.505 \\-2.064 & 3.165 & 5.723 & 3.898 & -5.551 & 4.059 & 1.257 & 0.465 & 14.236 & 2.017 \\2.037 & 1.737 & 1.516 & 0.256 & 0.831 & 0.182 & -1.529 & -2.505 & 2.017 & 6.648 \\
\end{bmatrix}.$$\\
\\
The second order term of the temporary transaction costs in Regime 1 is $tr_1^{a,\beta}= 10^{-4}*$
$$\begin{bmatrix}
5.490 & 0.563 & -1.166 & -0.140 & 1.593 & -0.253 & -1.849 & 2.440 & -1.562 & -0.320 \\0.563 & 8.075 & -1.394 & -1.178 & -1.079 & -4.500 & -0.618 & -3.111 & 0.423 & 0.545 \\-1.166 & -1.394 & 4.350 & 3.152 & 2.301 & -1.174 & 4.823 & 1.611 & -0.194 & -0.229 \\-0.140 & -1.178 & 3.152 & 11.844 & -1.752 & 0.593 & 5.164 & 1.859 & -2.080 & -0.495 \\1.593 & -1.079 & 2.301 & -1.752 & 8.880 & -3.342 & 2.145 & 0.396 & -4.176 & -0.738 \\-0.253 & -4.500 & -1.174 & 0.593 & -3.342 & 7.005 & -1.554 & 5.512 & -1.827 & 1.292 \\-1.849 & -0.618 & 4.823 & 5.164 & 2.145 & -1.554 & 6.202 & 2.014 & -1.021 & 0.413 \\2.440 & -3.111 & 1.611 & 1.859 & 0.396 & 5.512 & 2.014 & 21.046 & -3.422 & 6.376 \\-1.562 & 0.423 & -0.194 & -2.080 & -4.176 & -1.827 & -1.021 & -3.422 & 9.075 & -2.181 \\-0.320 & 0.545 & -0.229 & -0.495 & -0.738 & 1.292 & 0.413 & 6.376 & -2.181 & 9.014
\end{bmatrix}.$$\\
\\
The first order term of the temporary transaction costs in Regime 2 is $tr_2^{a,\alpha}= 10^{-3}*$
$$\begin{bmatrix}
19.690 & 9.386 & 4.315 & -0.168 & -7.628 & -5.980 & 2.741 & 2.121 & -2.718 & 0.379 \\9.386 & 11.563 & 2.139 & 0.219 & -3.470 & 0.108 & 1.488 & -0.067 & -7.138 & -1.483 \\4.315 & 2.139 & 7.931 & -1.516 & -0.559 & -6.497 & -2.678 & -1.220 & -1.249 & 0.507 \\-0.168 & 0.219 & -1.516 & 3.929 & -0.359 & 1.965 & 2.006 & -0.536 & -4.214 & 0.349 \\-7.628 & -3.470 & -0.559 & -0.359 & 13.445 & -1.178 & -2.841 & -2.770 & -0.291 & -0.866 \\-5.980 & 0.108 & -6.497 & 1.965 & -1.178 & 8.572 & 1.160 & 0.387 & -0.141 & -1.485 \\2.741 & 1.488 & -2.678 & 2.006 & -2.841 & 1.160 & 11.993 & 4.089 & -4.767 & 1.183 \\2.121 & -0.067 & -1.220 & -0.536 & -2.770 & 0.387 & 4.089 & 4.971 & -0.392 & -3.187 \\-2.718 & -7.138 & -1.249 & -4.214 & -0.291 & -0.141 & -4.767 & -0.392 & 14.481 & 2.968 \\0.379 & -1.483 & 0.507 & 0.349 & -0.866 & -1.485 & 1.183 & -3.187 & 2.968 & 11.152 \\
\end{bmatrix}.$$\\
\\
The second order term of the temporary transaction costs in Regime 2 is $tr_2^{a,\beta}= 10^{-4}*$
$$\begin{bmatrix}
9.690 & 2.709 & -3.930 & -3.289 & -0.784 & 1.399 & -0.552 & -1.165 & 9.026 & -1.613 \\2.709 & 3.239 & -0.721 & -0.498 & 1.071 & -2.102 & -1.108 & -1.505 & 4.107 & 0.759 \\-3.930 & -0.721 & 9.989 & 7.926 & 1.424 & 3.760 & -2.321 & -1.822 & -2.501 & 3.902 \\-3.289 & -0.498 & 7.926 & 17.919 & 4.141 & 8.807 & -0.557 & 2.269 & -3.158 & 2.063 \\-0.784 & 1.071 & 1.424 & 4.141 & 4.182 & 0.704 & -0.889 & 0.021 & 2.390 & 4.267 \\1.399 & -2.102 & 3.760 & 8.807 & 0.704 & 9.926 & 0.393 & 2.455 & 1.037 & -1.195 \\-0.552 & -1.108 & -2.321 & -0.557 & -0.889 & 0.393 & 3.034 & 0.530 & -3.759 & -2.570 \\-1.165 & -1.505 & -1.822 & 2.269 & 0.021 & 2.455 & 0.530 & 5.780 & -1.839 & -0.698 \\9.026 & 4.107 & -2.501 & -3.158 & 2.390 & 1.037 & -3.759 & -1.839 & 17.794 & 2.683 \\-1.613 & 0.759 & 3.902 & 2.063 & 4.267 & -1.195 & -2.570 & -0.698 & 2.683 & 8.690 \\
\end{bmatrix}.$$\\
\\
The first order of the permanent transaction costs under Regime 1 is
$tr_1^{b,\alpha}= 10^{-4}*$
$$
\begin{bmatrix}
5.035 & -0.551 & -0.541 & 1.025 & -0.035 & 2.531 & 0.335 & 2.401 & -2.769 & 0.918 \\-0.551 & 6.637 & 2.499 & -1.521 & -3.435 & -1.321 & 3.623 & -1.880 & -3.730 & 1.157 \\-0.541 & 2.499 & 7.291 & -0.946 & -2.768 & -1.219 & 1.068 & 0.123 & 0.632 & 0.931 \\1.025 & -1.521 & -0.946 & 4.408 & -0.198 & 0.944 & -1.923 & -0.940 & 2.068 & -0.841 \\-0.035 & -3.435 & -2.768 & -0.198 & 6.861 & -2.372 & -1.797 & 2.261 & 1.179 & -1.625 \\2.531 & -1.321 & -1.219 & 0.944 & -2.372 & 5.563 & 0.248 & 0.025 & 0.734 & 1.482 \\0.335 & 3.623 & 1.068 & -1.923 & -1.797 & 0.248 & 4.307 & -0.624 & -3.308 & 1.189 \\2.401 & -1.880 & 0.123 & -0.940 & 2.261 & 0.025 & -0.624 & 3.913 & -1.498 & -0.394 \\-2.769 & -3.730 & 0.632 & 2.068 & 1.179 & 0.734 & -3.308 & -1.498 & 8.184 & -0.705 \\0.918 & 1.157 & 0.931 & -0.841 & -1.625 & 1.482 & 1.189 & -0.394 & -0.705 & 1.572 \\
\end{bmatrix}.$$ \\
\\
The second order of the permanent transaction costs under Regime 1 is $tr_1^{b,\beta}= 10^{-4}*$
$$
\begin{bmatrix}
6.249 & 0.481 & -0.711 & -1.501 & 0.149 & 2.259 & -0.098 & -2.199 & -2.777 & 0.288 \\0.481 & 2.756 & -0.725 & 2.293 & -2.288 & -0.943 & 1.568 & -0.395 & -1.051 & 0.080 \\-0.711 & -0.725 & 3.683 & -1.936 & 0.863 & -1.560 & 0.742 & 0.203 & -1.103 & -0.210 \\-1.501 & 2.293 & -1.936 & 8.259 & -1.707 & -0.048 & 1.915 & 2.110 & 0.359 & 0.831 \\0.149 & -2.288 & 0.863 & -1.707 & 5.309 & 0.481 & -1.548 & 0.468 & 0.700 & -0.447 \\2.259 & -0.943 & -1.560 & -0.048 & 0.481 & 5.259 & -0.507 & -0.873 & -0.768 & 1.346 \\-0.098 & 1.568 & 0.742 & 1.915 & -1.548 & -0.507 & 3.005 & -0.861 & -1.248 & -0.275 \\-2.199 & -0.395 & 0.203 & 2.110 & 0.468 & -0.873 & -0.861 & 3.624 & 0.032 & 2.233 \\-2.777 & -1.051 & -1.103 & 0.359 & 0.700 & -0.768 & -1.248 & 0.032 & 3.611 & -1.906 \\0.288 & 0.080 & -0.210 & 0.831 & -0.447 & 1.346 & -0.275 & 2.233 & -1.906 & 3.343 \\
\end{bmatrix}.$$ \\
\\
The first order of the permanent transaction costs under Regime 2 is $tr_2^{b,\alpha}= 10^{-4}*$
$$\begin{bmatrix}
8.779 & 1.893 & -2.619 & 0.149 & 0.738 & 0.341 & -0.635 & -0.432 & 5.009 & 1.955 \\1.893 & 10.167 & -1.012 & 0.407 & -4.759 & 3.833 & 0.634 & 2.900 & -3.953 & -5.518 \\-2.619 & -1.012 & 5.218 & -0.531 & 0.416 & -1.977 & 2.596 & 1.945 & 0.156 & -1.806 \\0.149 & 0.407 & -0.531 & 4.638 & 1.838 & -1.752 & 0.068 & 2.911 & 0.319 & 2.052 \\0.738 & -4.759 & 0.416 & 1.838 & 9.351 & -5.021 & 1.158 & -5.298 & 9.184 & 3.063 \\0.341 & 3.833 & -1.977 & -1.752 & -5.021 & 8.866 & 1.430 & 1.172 & -3.768 & -3.164 \\-0.635 & 0.634 & 2.596 & 0.068 & 1.158 & 1.430 & 8.626 & -0.017 & 0.874 & -3.669 \\-0.432 & 2.900 & 1.945 & 2.911 & -5.298 & 1.172 & -0.017 & 12.769 & -8.050 & 3.116 \\5.009 & -3.953 & 0.156 & 0.319 & 9.184 & -3.768 & 0.874 & -8.050 & 24.148 & -1.431 \\1.955 & -5.518 & -1.806 & 2.052 & 3.063 & -3.164 & -3.669 & 3.116 & -1.431 & 9.862 \\
\end{bmatrix}.$$\\
\\
The second order of the permanent transaction costs under Regime 2 is $tr_2^{b,\beta}= 10^{-4}*$
$$
\begin{bmatrix}
6.926 & 2.159 & -2.374 & 0.424 & 1.113 & -5.058 & 3.678 & 1.643 & 0.229 & -5.776 \\2.159 & 9.391 & 2.152 & 1.701 & -1.702 & 0.799 & -0.421 & -3.545 & -4.785 & 0.266 \\-2.374 & 2.152 & 8.808 & -0.246 & -4.413 & -0.539 & -2.243 & -1.113 & 0.188 & 0.988 \\0.424 & 1.701 & -0.246 & 3.004 & -1.364 & -1.056 & -0.708 & -1.175 & -2.069 & 4.284 \\1.113 & -1.702 & -4.413 & -1.364 & 7.321 & -4.374 & 1.537 & -0.929 & 0.903 & -2.537 \\-5.058 & 0.799 & -0.539 & -1.056 & -4.374 & 14.364 & -2.294 & -0.409 & -2.362 & 2.270 \\3.678 & -0.421 & -2.243 & -0.708 & 1.537 & -2.294 & 5.195 & 0.422 & 0.739 & -7.302 \\1.643 & -3.545 & -1.113 & -1.175 & -0.929 & -0.409 & 0.422 & 8.022 & 4.320 & -1.812 \\0.229 & -4.785 & 0.188 & -2.069 & 0.903 & -2.362 & 0.739 & 4.320 & 5.977 & -0.875 \\-5.776 & 0.266 & 0.988 & 4.284 & -2.537 & 2.270 & -7.302 & -1.812 & -0.875 & 17.556 \\
\end{bmatrix}.$$ \\
\\
The return of the risky assets under Regime 1 follows a multi-variate normal distribution with mean $[-0.0096, -0.0018,  0.0087,  0.0069, -0.0035,0.0232, -0.0002 , -0.0068,  0.0146,  0.0183]^T$ and covariance matrix
$$10^{-4}*\begin{bmatrix}
1.303 & -0.548 & 0.255 & 0.286 & 0.305 & 0.177 & 0.004 & -0.129 & 0.102 & 0.241 \\-0.548 & 1.015 & -0.181 & -0.435 & -0.000 & -0.424 & -0.035 & 0.110 & -0.355 & -0.639 \\0.255 & -0.181 & 1.214 & -0.004 & 0.605 & 0.335 & 0.456 & 0.030 & -0.103 & 0.377 \\0.286 & -0.435 & -0.004 & 1.197 & -0.028 & 0.209 & 0.100 & -0.382 & 0.074 & 0.222 \\0.305 & -0.000 & 0.605 & -0.028 & 0.562 & -0.037 & 0.279 & -0.114 & -0.222 & 0.134 \\0.177 & -0.424 & 0.335 & 0.209 & -0.037 & 0.858 & -0.155 & 0.079 & 0.221 & 0.447 \\0.004 & -0.035 & 0.456 & 0.100 & 0.279 & -0.155 & 0.518 & -0.075 & -0.147 & 0.031 \\-0.129 & 0.110 & 0.030 & -0.382 & -0.114 & 0.079 & -0.075 & 0.267 & 0.061 & -0.098 \\0.102 & -0.355 & -0.103 & 0.074 & -0.222 & 0.221 & -0.147 & 0.061 & 0.504 & 0.386 \\0.241 & -0.639 & 0.377 & 0.222 & 0.134 & 0.447 & 0.031 & -0.098 & 0.386 & 1.065 \\
\end{bmatrix},$$ and that under Regime 2 has mean $[-0.0082, -0.0140,  0.0103, -0.0205, -0.0123,  0.0097, -0.0006,\\ -0.0026,  0.0035, -0.0015]^T$ and covariance matrix
$$10^{-4}*\begin{bmatrix}
1.025 & 0.012 & 0.323 & 0.288 & -0.077 & -0.167 & 0.290 & 0.354 & -0.135 & 0.643 \\0.012 & 0.767 & 0.093 & 0.303 & -0.038 & 0.150 & -0.185 & 0.262 & 0.058 & 0.584 \\0.323 & 0.093 & 0.800 & 0.344 & -0.001 & -0.269 & -0.238 & -0.051 & -0.330 & 0.329 \\0.288 & 0.303 & 0.344 & 1.648 & 0.078 & -0.112 & -0.190 & 0.551 & -0.252 & 0.601 \\-0.077 & -0.038 & -0.001 & 0.078 & 0.205 & 0.195 & 0.031 & 0.040 & 0.065 & -0.006 \\-0.167 & 0.150 & -0.269 & -0.112 & 0.195 & 0.837 & 0.116 & 0.292 & 0.057 & 0.055 \\0.290 & -0.185 & -0.238 & -0.190 & 0.031 & 0.116 & 0.468 & -0.106 & 0.281 & -0.017 \\0.354 & 0.262 & -0.051 & 0.551 & 0.040 & 0.292 & -0.106 & 0.978 & -0.128 & 0.518 \\-0.135 & 0.058 & -0.330 & -0.252 & 0.065 & 0.057 & 0.281 & -0.128 & 0.697 & 0.210 \\0.643 & 0.584 & 0.329 & 0.601 & -0.006 & 0.055 & -0.017 & 0.518 & 0.210 & 1.579
\end{bmatrix}.$$ \\
\\
The initial price of the ten assets are \$4.076, \$3.712, \$2.279, \$1.786, \$3.686, \$4.372, \$1.065, \$3.571, \$2.771 , and \$4.592, respectively. The regime transition matrix is $\begin{bmatrix}
0.95& 0.05\\
0.08& 0.92\\
\end{bmatrix}$.

\section{Example with one security}
In this section we present an example with sale of one security in 10 periods. To reduce the state space, we break the amount to be sold into 20 equal chunks. In this section, the relative risk aversion parameter $\gamma$ is assumed to be -2. We will provide three different scenarios.

\subsection{Scenario 1: One regime with high return, and one regime with low return.}\label{scenario1}
\begin{figure}[ht!]
\center
\begin{subfigure}{0.48\textwidth}
\includegraphics[width=0.9\linewidth]{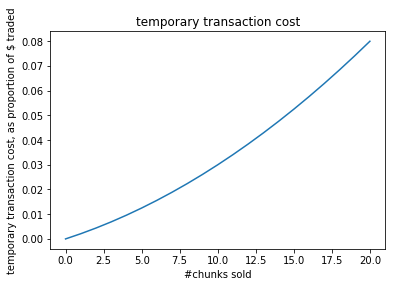} 
\caption{Temporary transaction costs versus number of chunks in an order.}
\label{fig:temp_tcost_1}
\end{subfigure}
\begin{subfigure}{0.48\textwidth}
\includegraphics[width=0.9\linewidth]{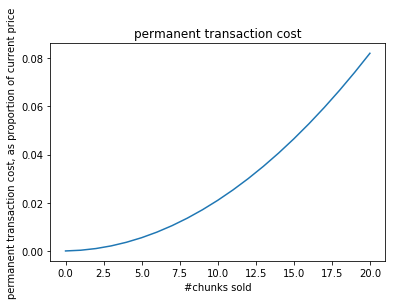}
\caption{Permanent transaction costs versus number of chunks in an order.}
\label{fig:perm_tcost_1}
\end{subfigure}
\caption{\textbf{Scenario 1.}}
\label{tcost_1}
\end{figure}

Assume there are two regimes in the market, one with high asset returns and the other with low returns. In the first regime, the return follows a normal distribution $\mathcal{N}(0.001, 0.00085^2)$ within one period time; in the second regime, the return follows $\mathcal{N}(-0.00008, 0.001^2)$ within one period time. The transition matrix of regimes is 
$\begin{bmatrix}
0.95 & 0.05\\
0.08 & 0.92
\end{bmatrix}$.\\
\\
Transaction costs under both regimes are taken to be the same in this scenario. The temporary transaction cost is $tr_1^a = tr_2^a = 2*10^{-3}x + 10^{-4}x^2$ for each \$1 traded, where $x$ is the number of traded chunks. The permanent transaction cost is $tr_1^b=tr_2^b= 10^{-4}x + 2*10^{-4}x^2$. The price immediately after the trade equals $(1-tr_i^b)$ times the price right before the trade. Both types of transaction costs are convex as Figures \ref{fig:temp_tcost_1} and \ref{fig:perm_tcost_1} show.

\begin{figure}[ht!]
\center
\begin{subfigure}{0.48\textwidth}
\includegraphics[width=0.9\linewidth]{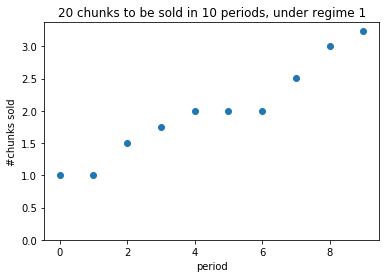} 
\caption{The number of chunks to sell in each period.}
\label{fig:sc1_exp1_x}
\end{subfigure}
\begin{subfigure}{0.48\textwidth}
\includegraphics[width=0.9\linewidth]{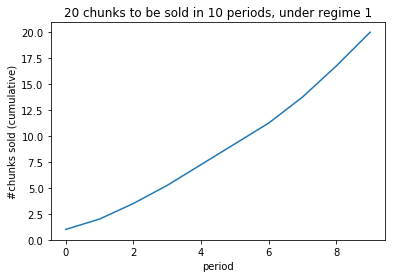}
\caption{Cumulative number of chunks sold.}
\label{fig:sc1_exp1_cum_x}
\end{subfigure}
\caption{\textbf{Scenario 1, Example 1.1.}}
\label{fig:sc1_exp1}
\end{figure}

\noindent
Below we provide a few sample paths under Scenario 1, and analyze how different realized regimes may alter the selling schedule.\\
\\
\textbf{Example 1.1 (under Regime 1 - high returns).} In this sample path, the market stays in Regime 1 throughout the horizon of interest. The amount of chunks to sell in each period and the cumulative chunks sold are plotted in Figure \ref{fig:sc1_exp1}. Since the permanent impact cost carries on and the price tend to increase under Regime 1, the trader benefits from holding some asset toward the end of the horizon. In general, under Regime 1, the amount to sell increases as the horizon approaches. Therefore the cumulative number of chunks sold tends to be convex.\\ 

\begin{figure}[ht!]
\center
\begin{subfigure}{0.48\textwidth}
\includegraphics[width=0.9\linewidth]{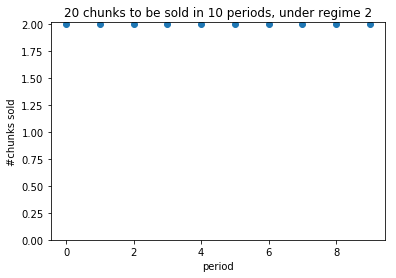} 
\caption{The number of chunks to sell in each period.}
\label{fig:sc1_exp2_x}
\end{subfigure}
\begin{subfigure}{0.48\textwidth}
\includegraphics[width=0.9\linewidth]{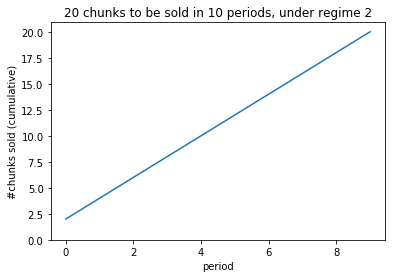}
\caption{Cumulative number of chunks sold.}
\label{fig:sc1_exp2_cum_x}
\end{subfigure}
\caption{\textbf{Scenario 1, Example 1.2.}}
\label{fig:sc1_exp2}
\end{figure}

\noindent
\textbf{Example 1.2 (under Regime 2 - low returns).} In this example, the market stays in Regime 2 throughout the horizon. Selling schedule appears in Figure \ref{fig:sc1_exp2}.
The price tends to decrease over time, so the agent would sell more chunks at the beginning, compared to the previous example. On the other hand, as both temporary and permanent impacts are convex, the trader also avoids selling the asset too fast. The cumulative number of chunks sold tend to be less convex (or more concave) than in Example 1.\\

\begin{figure}[ht!]
\center
\begin{subfigure}{0.9\textwidth}
\includegraphics[width=0.9\linewidth]{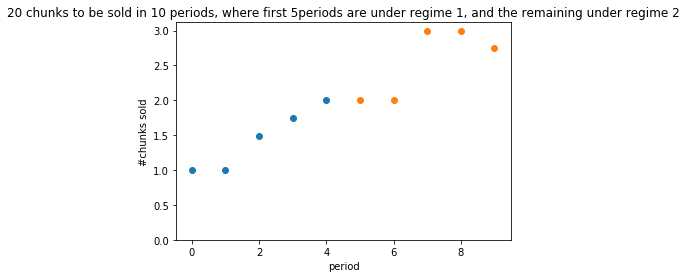} 
\caption{The number of chunks to sell in each period.}
\label{fig:sc1_exp3_x}
\end{subfigure}
\begin{subfigure}{0.9\textwidth}
\includegraphics[width=0.9\linewidth]{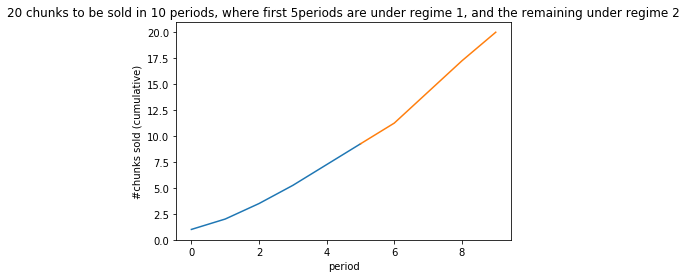}
\caption{Cumulative number of chunks sold.}
\label{fig:sc1_exp3_cum_x}
\end{subfigure}
\caption{\textbf{Scenario 1, Example 1.3.}}
\label{fig:sc1_exp3}
\end{figure}

\noindent
\textbf{Example 1.3 (mixed regime).} In this example, we take a sample path whose first 5 periods are under Regime 1, and last 5 periods are under Regime 2. The strategy appears in Figure \ref{fig:sc1_exp3}.
The selling schedule of first half is similar to that in Example 1, which is a result of non-anticipativity. Once the regime switches after the 5th period, the trader makes an effort to flatten the amount to sell, and the strategy in the latter half is similar to that in Example 2.\\

\subsection{Scenario 2: One regime with high liquidity, and one regime with low liquidity.}\label{scenario2}
Again we assume there are two regimes in the market, but this time, with the same return distribution and varying transaction costs under both regimes. In particular, the returns in both regimes follow $\mathcal{N}(0.001, 0.00085^2)$. The temporary and permanent transaction costs under both regimes are convex functions of number of traded chunks, while the transaction costs in Regime 2 is three times of that in Regime 1 (Figure \ref{}). This scenario captures the features that the liquidity of assets may vary from time to time. The transition matrix of regimes is 
$\begin{bmatrix}
0.95 & 0.05\\
0.08 & 0.92
\end{bmatrix}$.\\
\\
\begin{figure}[ht!]
\center
\begin{subfigure}{0.48\textwidth}
\includegraphics[width=0.9\linewidth]{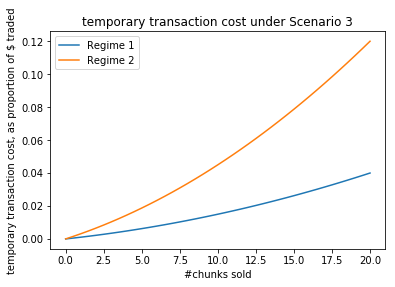} 
\caption{Temporary transaction costs versus number of chunks in an order.}
\label{fig:temp_tcost_2}
\end{subfigure}
\begin{subfigure}{0.48\textwidth}
\includegraphics[width=0.9\linewidth]{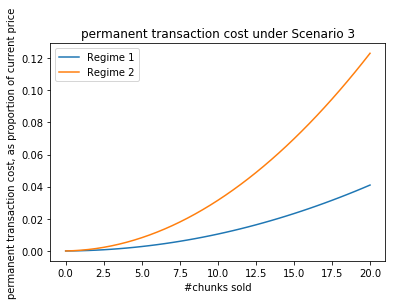}
\caption{Permanent transaction costs versus number of chunks in an order.}
\label{fig:perm_tcost_2}
\end{subfigure}
\caption{\textbf{Scenario 2.}}
\label{tcost_2}
\end{figure}

\begin{figure}[ht!]
\center
\begin{subfigure}{0.48\textwidth}
\includegraphics[width=0.9\linewidth]{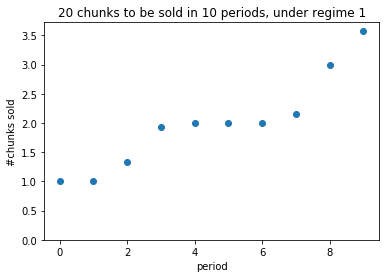} 
\caption{The number of chunks to sell in each period.}
\label{fig:sc2_exp1_x}
\end{subfigure}
\begin{subfigure}{0.48\textwidth}
\includegraphics[width=0.9\linewidth]{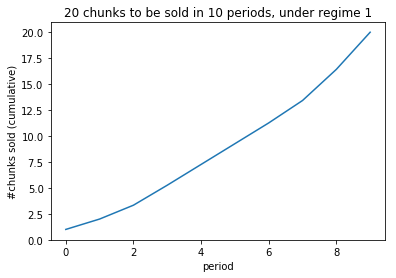}
\caption{Cumulative number of chunks sold.}
\label{fig:sc2_exp1_cum_x}
\end{subfigure}
\caption{\textbf{Scenario 2, Example 2.1.}}
\label{fig:sc2_exp1}
\end{figure}

\noindent
\textbf{Example 2.1 (under Regime 1 - high liquidity).} In this sample path, the market stays in Regime 1 throughout the horizon of interest. The trading scheme appears in Figure \ref{fig:sc2_exp1}. Since the transaction costs are lower in this regime, trading a large volume would hurt the price much less than in the other regime. As a result, the trader can sell relatively heavily at the beginning when she starts with high liquidity, which provides substantial advantage in case the regime switches towards the end of horizon.\\

\begin{figure}[ht!]
\center
\begin{subfigure}{0.48\textwidth}
\includegraphics[width=0.9\linewidth]{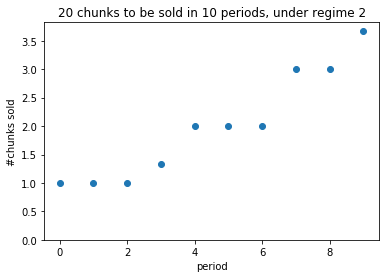} 
\caption{The number of chunks to sell in each period.}
\label{fig:sc2_exp2_x}
\end{subfigure}
\begin{subfigure}{0.48\textwidth}
\includegraphics[width=0.9\linewidth]{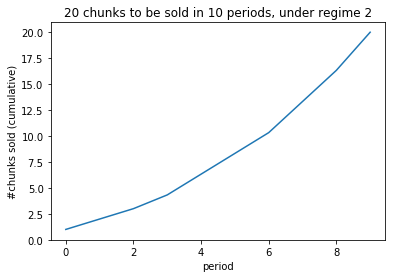}
\caption{Cumulative number of chunks sold.}
\label{fig:sc2_exp2_cum_x}
\end{subfigure}
\caption{\textbf{Scenario 2, Example 2.2.}}
\label{fig:sc2_exp2}
\end{figure}

\noindent
\textbf{Example 2.2 (under Regime 2 - low liquidity).} In this example, the market is illiquid throughout the horizon. Selling schedule appears in Figure \ref{fig:sc2_exp2}. With high transaction costs, the trader faces the trade-off between holding the asset to sell at a better mid price, and spread the sale evenly through the horizon to avoid price drops due to permanent transaction costs. \\

\begin{figure}[ht!]
\center
\begin{subfigure}{0.9\textwidth}
\includegraphics[width=0.9\linewidth]{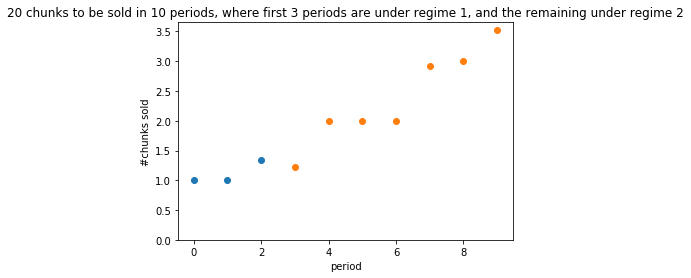} 
\caption{The number of chunks to sell in each period.}
\label{fig:sc2_exp3_x}
\end{subfigure}
\begin{subfigure}{0.9\textwidth}
\includegraphics[width=0.9\linewidth]{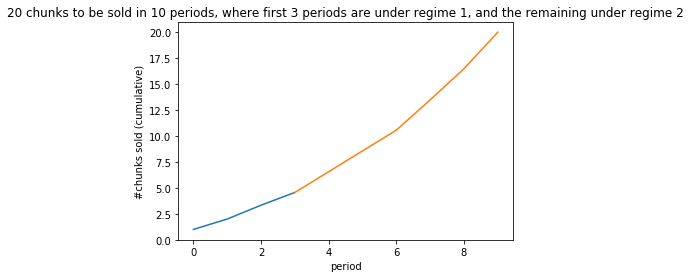}
\caption{Cumulative number of chunks sold.}
\label{fig:sc2_exp3_cum_x}
\end{subfigure}
\caption{\textbf{Scenario 2, Example 2.3.}}
\label{fig:sc2_exp3}
\end{figure}

\noindent
\textbf{Example 2.3 (mixed regime).} In this example, we take a sample path whose first 3 periods are under Regime 1, and last 7 periods are under Regime 2. The strategy appears in Figure \ref{fig:sc2_exp3}. With the non-anticipativity constraints, the selling schedule in the first 3 periods is the same as in Example 2.1. After the regime switches to low liquidity, the strategy becomes similar to that in Example 2.2.\\

\section{Sensitivity Analysis with one security}
In this section, we carry out sensitivity analysis of how changes in transition probabilities and/or transaction costs would impact the trading schedule. Intuitively, the more likely regimes switch, the more similar selling patterns under different realized regime paths, as there are less information that can be predicted from the current regime. The more distinct transaction costs under different regimes, the more distinguishes in the patterns in varying realized regime paths. We illustrate the idea with quantitative examples.\\

\subsection{Sensitivity analysis on transition probabilities}
In this subsection, we alter the transition probabilities based on the Scenario \ref{scenario1}. The return patterns and transaction costs under both regimes are kept the same as in Scenario \ref{scenario1}.\\

\subsubsection{Increase the likelihood of regime switching}
We investigate the change in selling strategy when the likelihood of regime switching increases, i.e., the regimes are less stable than in Scenario \ref{scenario1}. In such cases, knowing the current regime has decreasing power of predicting future market dynamics, and thus the selling schedule is expected to be more similar under different realized regime paths. In particular, we present strategies when the transition matrix is 
$\begin{bmatrix}
0.9 & 0.1\\
0.16 & 0.84
\end{bmatrix}$,
$\begin{bmatrix}
0.8 & 0.2 \\
0.32 & 0.68
\end{bmatrix}$ and 
$\begin{bmatrix}
0.5 & 0.5\\
0.5 & 0.5
\end{bmatrix}$.\\

\subsubsection{Decrease the likelihood of regime switching}
Now we decrease the likelihood of regime switching, and study the selling schedule under the extreme case where the transition matrix is 
$\begin{bmatrix}
1 & 0\\
0 & 1
\end{bmatrix}$, i.e., throughout the horizon of interest, the market stays in the regime that it starts with.\\

\subsubsection{Trading schedule under varying transition probabilities}
The trading strategy under several realized paths are included in Table \ref{tbl:sensitivity_tm}. We observe that the less frequent the regimes may switch, the more distinguish trading strategies we have under different realized paths. If the regimes is stable and does not change, one is able to optimize the sale over a single-regime framework; on the other hand, if current regime provides zero predicting power of future regimes, the trading strategy are alike under any realized paths due to the non-anticipativity constraints.\\

\begin{table}[ht!]
  \centering
  \begin{tabular}{ | m{0.25\textwidth} | m{0.35\textwidth} | m{0.35\textwidth} | }
    \hline
    Transition matrix between the regimes & A realized path under Regime 1 throughout the horizon & A realized path under Regime 2 throughout the horizon \\ \hline
    $\begin{bmatrix}
    1 & 0\\
    0 & 1
    \end{bmatrix}$
    &
    \begin{minipage}{.3\textwidth}
      \includegraphics[width=\linewidth]{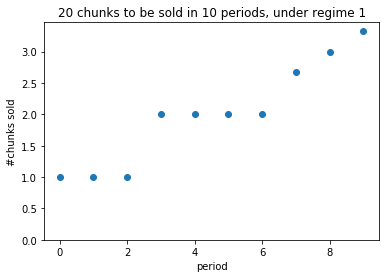}
    \end{minipage}
    & 
    \begin{minipage}{.3\textwidth}
      \includegraphics[width=\linewidth]{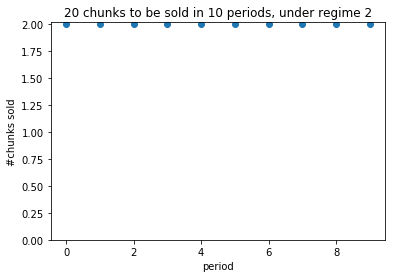}
    \end{minipage}
    \\ \hline
    $\begin{bmatrix}
    0.95 & 0.05\\
    0.08 & 0.92
    \end{bmatrix}$
    &
    \begin{minipage}{.3\textwidth}
      \includegraphics[width=\linewidth]{sell_regime1.png}
    \end{minipage}
    & 
    \begin{minipage}{.3\textwidth}
      \includegraphics[width=\linewidth]{sell_regime2.png}
    \end{minipage}
    \\ \hline
    $\begin{bmatrix}
    0.90 & 0.10\\
    0.16 & 0.84
    \end{bmatrix}$
    &
    \begin{minipage}{.3\textwidth}
      \includegraphics[width=\linewidth]{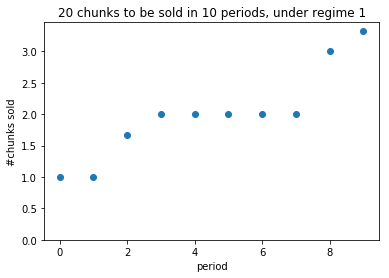}
    \end{minipage}
    & 
    \begin{minipage}{.3\textwidth}
      \includegraphics[width=\linewidth]{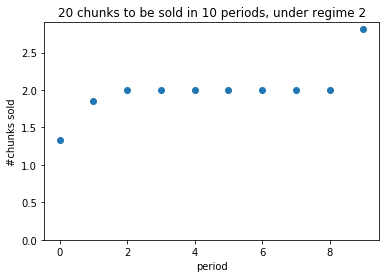}
    \end{minipage}
    \\ \hline
    $\begin{bmatrix}
    0.80 & 0.20\\
    0.32 & 0.68
    \end{bmatrix}$
    &
    \begin{minipage}{.3\textwidth}
      \includegraphics[width=\linewidth]{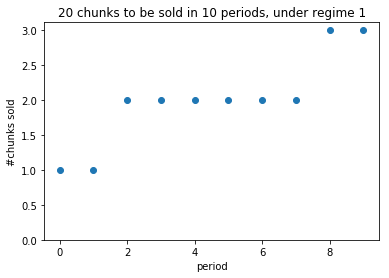}
    \end{minipage}
    & 
    \begin{minipage}{.3\textwidth}
      \includegraphics[width=\linewidth]{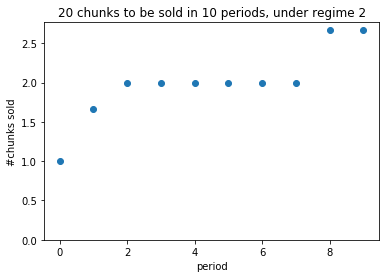}
    \end{minipage}
    \\ \hline
    $\begin{bmatrix}
    0.50 & 0.50\\
    0.50 & 0.50
    \end{bmatrix}$
    &
    \begin{minipage}{.3\textwidth}
      \includegraphics[width=\linewidth]{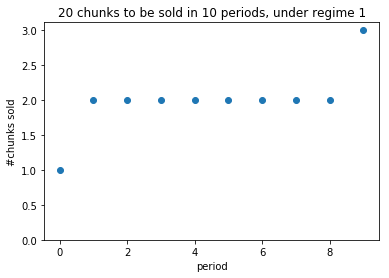}
    \end{minipage}
    & 
    \begin{minipage}{.3\textwidth}
      \includegraphics[width=\linewidth]{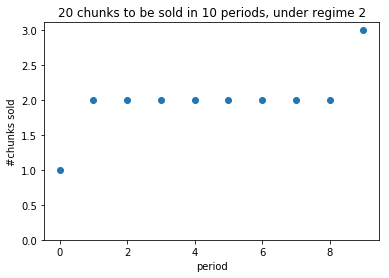}
    \end{minipage}
    \\ \hline
  \end{tabular}
  \caption{Trading schedule with different transition probabilities between two regimes}\label{tbl:sensitivity_tm}
\end{table}

\subsection{Sensitivity analysis on transaction costs}
We vary the transaction cost in Regime 2 based on the Scenario \ref{scenario2}. In Scenario \ref{scenario2}, both temporary and permanent transaction cost in Regime 2 is three times of that in Regime 1. In this subsection, we will keep the transaction costs under Regime 1, and alter the transaction costs under Regime 2. The return distributions are kept the same as in Scenario \ref{scenario2}, and the transition matrix is unchanged.\\

\subsubsection{Increase transaction cost under Regime 2}
\begin{figure}[ht!]
\center
\begin{subfigure}{0.48\textwidth}
\includegraphics[width=0.9\linewidth]{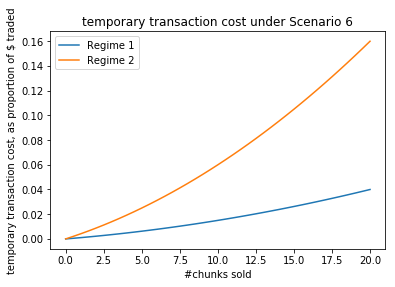} 
\caption{Temporary transaction costs versus number of chunks in an order.}
\label{fig:temp_tcost_increased}
\end{subfigure}
\begin{subfigure}{0.48\textwidth}
\includegraphics[width=0.9\linewidth]{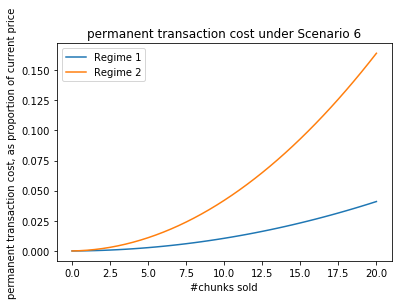}
\caption{Permanent transaction costs versus number of chunks in an order.}
\label{fig:perm_tcost_increased}
\end{subfigure}
\caption{Transaction costs in Regime 2 is four times of that in Regime 1.}
\label{tcost_increased}
\end{figure}

Now assume the transaction costs under Regime 2 is four times of that under Regime 1, as plotted in Figure \ref{tcost_increased}.

\subsubsection{Decrease transaction cost under Regime 2}
\begin{figure}[ht!]
\center
\begin{subfigure}{0.48\textwidth}
\includegraphics[width=0.9\linewidth]{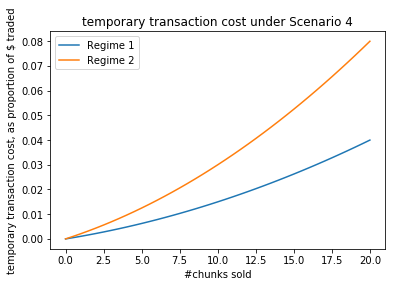} 
\caption{Temporary transaction costs versus number of chunks in an order.}
\label{fig:temp_tcost_increased1}
\end{subfigure}
\begin{subfigure}{0.48\textwidth}
\includegraphics[width=0.9\linewidth]{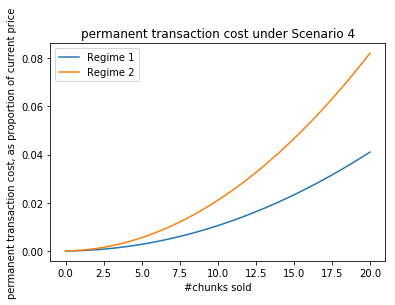}
\caption{Permanent transaction costs versus number of chunks in an order.}
\label{fig:perm_tcost_increased1}
\end{subfigure}
\caption{Transaction costs in Regime 2 is two times of that in Regime 1.}
\label{tcost_decreased1}
\end{figure}

\begin{figure}[ht!]
\center
\begin{subfigure}{0.48\textwidth}
\includegraphics[width=0.9\linewidth]{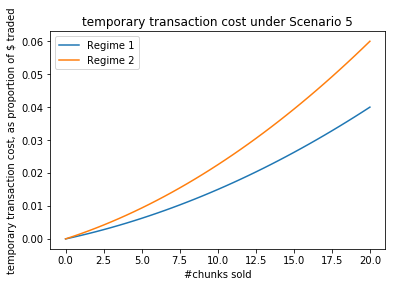} 
\caption{Temporary transaction costs versus number of chunks in an order.}
\label{fig:temp_tcost_increased2}
\end{subfigure}
\begin{subfigure}{0.48\textwidth}
\includegraphics[width=0.9\linewidth]{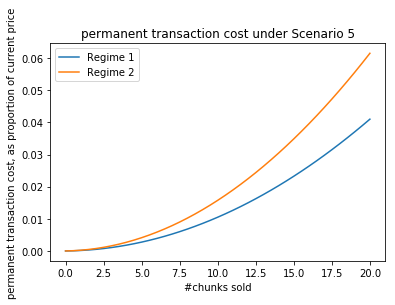}
\caption{Permanent transaction costs versus number of chunks in an order.}
\label{fig:perm_tcost_increased2}
\end{subfigure}
\caption{Transaction costs in Regime 2 is 1.5 times of that in Regime 1.}
\label{tcost_decreased2}
\end{figure}

Now assume the transaction costs under Regime 2 is two times or 1.5 times of that under Regime 1, as plotted in Figure \ref{tcost_decreased1} and Figure \ref{tcost_decreased2}.

\subsubsection{Trading schedule under varying transaction costs}
The trading strategy under several realized paths are included in Table \ref{tbl:sensitivity_tcost}. When the difference between transaction costs under distinct regimes decreases, the strategy under various realized paths converges. As Regime 2 features higher transaction costs, it is observed that an agent tends to sell the asset earlier rather than later if she is currently under Regime 1. This is not surprising, as the agent takes advantage of current regime to avoid paying unreasonably high transaction fees in case the market switches to the illiquid regime.

\begin{table}[ht!]
  \centering
  \begin{tabular}{ | m{0.3\textwidth} | m{0.33\textwidth} | m{0.33\textwidth} | }
    \hline
    Transaction cost & A realized path under Regime 1 throughout the horizon & A realized path under Regime 2 throughout the horizon \\ \hline
    Transaction costs in Regime 2 is \textbf{1.5 times} of that in Regime 1
    &
    \begin{minipage}{.3\textwidth}
      \includegraphics[width=\linewidth]{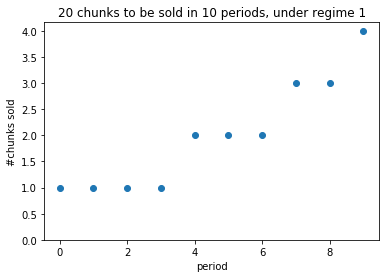}
    \end{minipage}
    & 
    \begin{minipage}{.3\textwidth}
      \includegraphics[width=\linewidth]{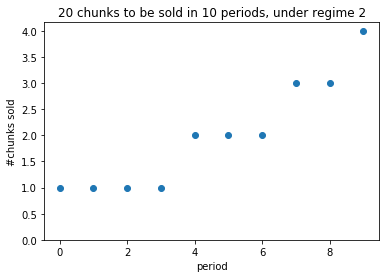}
    \end{minipage}
    \\ \hline
    Transaction costs in Regime 2 is \textbf{two times} of that in Regime 1
    &
    \begin{minipage}{.3\textwidth}
      \includegraphics[width=\linewidth]{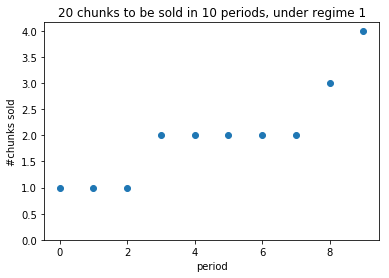}
    \end{minipage}
    & 
    \begin{minipage}{.3\textwidth}
      \includegraphics[width=\linewidth]{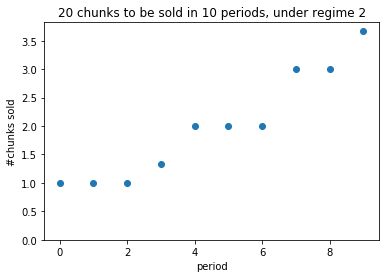}
    \end{minipage}
    \\ \hline
    Transaction costs in Regime 2 is \textbf{three times} of that in Regime 1
    &
    \begin{minipage}{.3\textwidth}
      \includegraphics[width=\linewidth]{sell_sc3_regime1.png}
    \end{minipage}
    & 
    \begin{minipage}{.3\textwidth}
      \includegraphics[width=\linewidth]{sell_sc3_regime2.png}
    \end{minipage}
    \\ \hline
    Transaction costs in Regime 2 is \textbf{four times} of that in Regime 1
    &
    \begin{minipage}{.3\textwidth}
      \includegraphics[width=\linewidth]{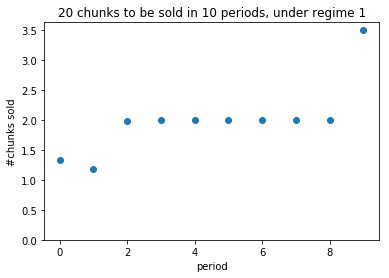}
    \end{minipage}
    & 
    \begin{minipage}{.3\textwidth}
      \includegraphics[width=\linewidth]{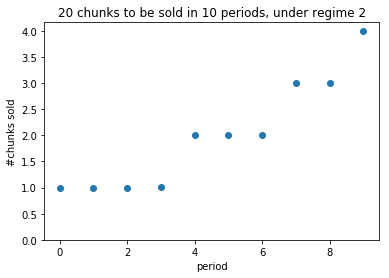}
    \end{minipage}
    \\ \hline
  \end{tabular}
  \caption{Trading schedule under different transaction costs}\label{tbl:sensitivity_tcost}
\end{table}

\end{document}